\documentclass[preprint,prd,aps,showpacs,showkeys,nofootinbib]{revtex4}
\usepackage{graphicx}
\usepackage{dcolumn}
\usepackage{bm}
\usepackage{color}
\usepackage[dvipsnames]{xcolor}
\usepackage{amssymb}

\definecolor{light-gray}{gray}{0.78}
\definecolor{mid-gray}{gray}{0.55}
\definecolor{dark-gray}{gray}{0.32}

\begin{document}
\title{Study of $\tau\rightarrow e M^+ M^-$ decays in the N-B-LSSM}
\author{Rong-Zhi Sun$^{1,2,3}$, Shu-Min Zhao$^{1,2,3}$\footnote{zhaosm@hbu.edu.cn}, Shuang Di$^{1,2,3}$, Xing-Xing Dong$^{1,2,3,4}$\footnote{dongxx@hbu.edu.cn}}
\affiliation{$^1$ Department of Physics, Hebei University, Baoding 071002, China}
\affiliation{$^2$ Hebei Key Laboratory of High-precision Computation and Application of Quantum Field Theory, Baoding, 071002, China}
\affiliation{$^3$ Hebei Research Center of the Basic Discipline for Computational Physics, Baoding, 071002, China}
\affiliation{$^4$ Departamento de Fisica and CFTP, Instituto Superior T$\acute{e}$cnico, Universidade de Lisboa,
Av.Rovisco Pais 1,1049-001 Lisboa, Portugal}
\date{\today}

\begin{abstract}

Within the framework of the next to the minimal supersymmetric (SUSY) extension of the Standard Model (SM) with a local B-L gauge symmetry (N-B-LSSM), we study lepton flavor violating (LFV) $\tau\rightarrow e M^+ M^-$ decays: $\tau \rightarrow e \pi^+\pi^-$,~$\tau \rightarrow e \pi^+K^-$,~$\tau \rightarrow e K^+K^-$. According to the latest experimental data, the influence of different sensitive parameters on the branching ratios is considered. It can be seen from the numerical analysis that the main sensitive parameters and LFV sources are non-diagonal elements corresponding to the initial and final leptons. This work can provide a basis for discovering the existence of new physics (NP).

\end{abstract}
\keywords{beyond Standard Model, supersymmetry, lepton flavor violation, new physics.}
\maketitle
\section{Introduction}

As the cornerstone of particle physics, the SM has achieved great success with the detected lightest CP-even Higgs \cite{Weinberg:1967tq,Salam:1968rm,ATLAS:2012yve,CMS:2012qbp}. Nevertheless, the SM faces several critical limitations in explaining fundamental phenomena: First of all, due to the absence of right-handed neutrinos and the presence of only the Yukawa couplings of quarks and charged leptons, the SM predicts that neutrinos are strictly massless, which directly contradicts the neutrino oscillations revealed by Super-Kamioka Neutrino Detection Experiment (1998) \cite{Super-Kamiokande:1998kpq,neutrino1,neutrino2,neutrino3}. Secondly, no stable and non-electric new particles in the SM can explain the large number of dark matter components observed in the universe. In addition, there is also a gauge hierarchy problem in the SM, which describes the large difference between the weak energy scale $m_{EW}$ and the Planck energy scale $M_{Pl}$. Moreover, due to the extremely small neutrino mixing angle and the existence of Glashow-Iliopoulos-Maiani (GIM) mechanism, LFV processes in the SM are very tiny, which is far lower than the detection sensitivity of Belle II experiment. Furthermore, the SM does not unify the gravitational interaction. The Minimal Supersymmetric Standard Model (MSSM) is the minimal expansion of SUSY theory to the SM \cite{n0,n1,n2}. It can explain the hierarchy problem, ensure high-energy scale unification of the gauge coupling constants, and also provide candidates for dark matter, partially compensating for the deficiencies of the SM. However, the MSSM has not yet solved the $\mu$ problem and neutrino mass problem.

Building on the MSSM, the N-B-LSSM extends the gauge group to $SU(3)_C\otimes SU(2)_L \otimes U(1)_Y\otimes U(1)_{B-L}$, where B and L denote baryon and lepton numbers, respectively (as first proposed in Ref. \cite{han}). In this model, two Higgs singlets carrying opposite $B-L$ charges $\hat{\chi}_1$ and $\hat{\chi}_2$ are introduced to trigger the spontaneous breaking of the $U(1)_{B-L}$ symmetry; meanwhile, three generation right-handed neutrinos $\hat{\nu}_i$ acquire Majorana masses through coupling with $\hat{\chi}_1$, and the additional singlet $\hat{S}$ is used to solve the $\mu$ problem in the superpotential. Furthermore, under the N-B-LSSM model, lepton number violation and baryon number violation processes can also occur, which help to explain the asymmetry of matter-antimatter in the universe. Right-handed neutrinos generate tiny masses through the Type-I seesaw mechanism, consistent with neutrino oscillation experiments, and neutralinos (as the lightest MSSM particles) can exist as dark matter candidates. The superpotential includes a trilinear term $\lambda\hat{S}\hat{H}_u\hat{H}_d$; when $\hat{S}$ acquires a vacuum expectation value (VEV) $\frac{v_S}{\sqrt{2}}$, it induces an effective $\mu$ term $\mu=\lambda \frac{v_S}{\sqrt{2}}$, thereby naturally alleviating the $\mu$ problem of the MSSM. The enlarged Higgs sector extends the neutral CP-even mass matrix to 5$\times$5, offering greater flexibility to fit the observed 125.20 $\pm$ 0.11~GeV Higgs mass and predicting additional scalar states for future experimental exploration. Additionally, the N-B-LSSM extends the gauge symmetry by introducing an extra $U(1)_{B-L}$ gauge group and its corresponding gauge boson $B^{\prime\mu}$, along with two new gauge coupling constants $g_B$ and $g_{YB}$. The associated gaugino $\tilde{B}^\prime$, together with the Higgsinos $\tilde{\chi}_1$, $\tilde{\chi}_2$ and $\tilde{S}$, enlarge the neutralino mass matrix from $4 \times 4$ to $8 \times 8$. The introduction of right-handed neutrinos also doubles the dimension of the sneutrino mass matrix from $3 \times 3$ to $6 \times 6$, significantly enriching the flavor structure. These extensions allow sizable LFV signals to arise even with mild flavor-mixing parameters, thereby enhancing the model's predictive power in LFV processes. Moreover, R-parity is automatically conserved in the N-B-LSSM due to the extended gauge structure, defined by the relation $R_p = (-1)^{3(B-L)+2S}$, without requiring additional assumptions. Finally, high scale VEVs ($v_{\eta}$, $v_{\bar{\eta}}$ and $v_S$) alleviate the little hierarchy problem by reducing the dependence on electroweak fine-tuning.

In contrast, the next-to-minimal supersymmetric standard model
(NMSSM) introduces only one additional Higgs singlet superfield $\hat{S}$, with superpotential terms such as $\lambda \hat{H}_u \cdot \hat{H}_d \hat{S}$ and $\frac{1}{3} \kappa \hat{S}^3$, which also address the $\mu$ problem to some extent. However, it lacks neutrino mass generation mechanisms present in the N-B-LSSM. As a result, its phenomenological impact on LFV processes remains similar to that of the MSSM. In summary, the choice to study LFV processes within the N-B-LSSM framework is grounded in the model's multiple theoretical advantages, including the generation of neutrino masses, alleviation of the $\mu$ problem, a more flexible Higgs sector, automatic R-parity conservation, a richer particle and flavor structure. These features make the N-B-LSSM a more compelling platform for exploring LFV.

In the SM framework, the pion and kaon mesons play a critical role as pseudo-Goldstone bosons in low-energy quantum chromodynamics (QCD), being composed of a quark and an antiquark. The $\pi^+$ meson is constituted by an up quark ($u$) and an anti-down quark ($\bar{d}$), while the $\pi^-$ meson is formed by a down quark ($d$) and an anti-up quark ($\bar{u}$). These two mesons belong to an isospin triplet, embodying the effective degrees of freedom emerging from the spontaneous breaking of chiral symmetry. In contrast, owing to the inclusion of a strange quark, the structure of the kaon mesons is more distinct: the $K^+$ meson is composed of an up quark ($u$) and an anti-strange quark ($\bar{s}$), while the $K^-$ meson consists of a strange quark ($s$) and an anti-up quark ($\bar{u}$). As a consequence, the mass of the kaon is significantly higher than that of the pion, reflecting the effects of SU(3) flavor symmetry and its explicit breaking. In the N-B-LSSM, the processes in which a $\tau$ lepton decays to an electron plus a meson pair ($\tau \rightarrow e \pi^+\pi^-$, $\tau\rightarrow e\pi^+K^-$, $\tau \rightarrow e K^+ K^-$) provide a sensitive platform for probing new physics (NP) effects.

Over the past few decades, LFV has become one of the core directions in exploring NP beyond Standard Model (BSM). Due to its largest mass and rich decay channels, $\tau$ lepton shows its unique advantages in LFV searches.
Ref. \cite{Chen:2006hp} investigates LFV in $\tau$ decays within a SUSY seesaw model. It reveals that scalar-mediated $\tau\rightarrow\ell f_0(980)$ and $\tau\rightarrow\ell K^+K^-$ branching ratios can reach $\mathcal{O}(10^{-7})$, surpassing pseudoscalar channels $\tau\to\ell\eta^{(')}$. Moreover, it links $\tau\to\ell\mu^+\mu^-$ to these processes, thereby identifying critical experimental targets for probing scalar-mediated LFV mechanisms.
Ref. \cite{Arganda:2008jj} studies semileptonic LFV $\tau$ decays $\tau\rightarrow \mu PP,~\tau\rightarrow \mu P,~\tau\rightarrow \mu V$ in CMSSM-seesaw and NUHM-seesaw frameworks via full one-loop analysis of $\gamma$-,~Z- and Higgs-mediated contributions. It identifies discrepancies in predicted branching ratios for $\tau\rightarrow\mu\eta,~\tau\rightarrow\mu\eta'$ and $\tau\rightarrow\mu K^+K^-$, proposing these channels as critical tests for LFV and Higgs-sector dynamics, with simplified formulas to aid experimental validation.
Ref. \cite{Arhrib:2009xf} explores LFV $\tau\rightarrow \ell P(V)$ and $\tau\rightarrow 3\ell$ decays within the Type-III seesaw model. By constraining the parameter space via experimental limits from leptonic Z-boson decays, the study predicts branching ratios for these processes that aligns with current experimental upper bound.
Ref. \cite{Celis:2013xja} investigates LFV Higgs decays $h\rightarrow \tau \ell~(\ell=e,~\mu)$ and their connection to hadronic $\tau$-decays (e.g., $\tau\rightarrow \ell \pi\pi,~\tau\rightarrow \ell\eta^{(')}$), aiming to distinguish scalar and pseudoscalar couplings in the Higgs sectors through low-energy processes, while improving the theoretical description of relevant hadronic matrix elements.
Ref. \cite{Feruglio:2016gvd} discusses the violation of lepton flavour universality (LFU) in B-decays by incorporating quantum corrections, such as renormalization group equation (RGE) running from a high-energy scale $\Lambda$, and predicts potential signals in LFV processes like $\tau\rightarrow\mu \ell\ell,~\tau\rightarrow \mu \rho,~\tau\rightarrow \mu \pi,~\tau\rightarrow \mu\eta^{(')}$.
Ref. \cite{Konno:2020tmf} emphasizes that the Belle experiment has set upper limits on the branching ratios of $\tau$ LFV and lepton number violation (LNV) decays, and anticipates Belle II to further probe these $\tau$ LFV/LNV decays in the coming decades. With a 50-fold increase in statistics, Belle II may reach branching ratios of $\mathcal{O}(10^{-7})-\mathcal{O}(10^{-9})$ predicted by NP models, revealing possible signals of BSM.
Ref. \cite{Wang:2021nqr} studies LFV decays $\tau\rightarrow Pl (P=\pi,~\eta,~\eta';~l=\mu,~e)$ in the $U(1)_X$SSM. By analyzing the impact of sensitive parameters using the latest experimental data on $\tau\rightarrow Pe$ and $\tau\rightarrow P\mu$, the study identifies non-diagonal elements as the primary sources of LFV, providing a theoretical foundation for exploring NP.

We investigate the LFV processes of $\tau$ to electron and meson pairs within the framework of the N-B-LSSM model. Under the premise of fully considering the experimental limit of $\tau\rightarrow e \gamma$ process \cite{SRZ}, we derive the relevant Feynman diagrams and amplitudes, and conduct the detailed numerical analysis on the branching ratio of each process. During the analysis, the contributions of a variety of SUSY particles in the loop diagrams are considered separately, and the changing trends of various contributions in different parameter ranges are studied. The effects of different parameters on the branching ratios are shown through the graphical results, the feasible parameter regions to satisfy the experimental limits are identified, and the key parameters that have the greater impact on the results are analysed. The latest upper limits on the LFV branching ratios of $\tau \rightarrow e \pi^+\pi^-$, $\tau\rightarrow e\pi^+K^-$ and $\tau \rightarrow e K^+ K^-$ at 90\% confidence level (C.L.) \cite{pdg} are:
\begin{eqnarray}
&&{\rm BR}(\tau \rightarrow e \pi^+\pi^-)<2.3\times10^{-8},~~~{\rm BR}(\tau\rightarrow e\pi^+K^-)<3.7\times10^{-8},\nonumber\\
&&{\rm BR}(\tau \rightarrow e K^+ K^-)<3.4\times10^{-8}.
\end{eqnarray}

The paper is organized as follows. In Sec.II, we introduce the main content of N-B-LSSM, presenting the required mass matrices and corresponding couplings. Sec.III derives analytical formula for the branching ratios of the LFV processes $\tau \rightarrow e \pi^+\pi^-$, $\tau\rightarrow e\pi^+K^-$ and $\tau \rightarrow e K^+ K^-$. In Sec.IV, we determine the input parameters and perform the numerical analysis. Sec.V summarizes the conclusion of this study. Finally, some specific forms of Wilson coefficients that we need are collected in the appendix \ref{A1}.

\section{The main content of N-B-LSSM}

The N-B-LSSM extends the local gauge group of the MSSM to $SU(3)_C\otimes SU(2)_L \otimes U(1)_Y\otimes U(1)_{B-L}$. N-B-LSSM has new superfields beyond MSSM, including right-handed neutrinos $\hat{\nu}_i$ and three Higgs singlets $\hat{\chi}_1$,~$\hat{\chi}_2$,~$\hat{S}$. Through the Type-I seesaw mechanism, the light neutrinos obtain tiny mass at the tree level. Meanwhile, in the Higgs scalar part, the neutral CP-even components from $H_u$, $H_d$, $\chi_1$, $\chi_2$ and $S$ are mixed to form a $5\times5$ mass squared matrix. By combining the loop corrections of SUSY particles, the lightest CP-even Higgs mass can be modified to 125.20 $\pm$ 0.11~GeV \cite{LCTHiggs1,LCTHiggs2}. Furthermore, the sneutrinos are dispersed into CP-even sneutrinos and CP-odd sneutrinos, and their mass squared matrices are both extended to $6\times6$.

The superpotential in the N-B-LSSM is expressed as:
\begin{eqnarray}
&&W=-Y_d\hat{d}\hat{q}\hat{H}_d-Y_e\hat{e}\hat{l}\hat{H}_d-\lambda_2\hat{S}\hat{\chi}_1\hat{\chi}_2+\lambda\hat{S}\hat{H}_u\hat{H}_d\nonumber\\&&~~~~~~~+\frac{\kappa}{3}\hat{S}\hat{S}\hat{S}+Y_u\hat{u}\hat{q}\hat{H}_u+Y_{\chi}\hat{\nu}\hat{\chi}_1\hat{\nu}
+Y_\nu\hat{\nu}\hat{l}\hat{H}_u.
\end{eqnarray}

The explicit forms of the two Higgs doublets are as follows:
\begin{eqnarray}
&&H_{u}=\left(\begin{array}{c}H_{u}^+\\{1\over\sqrt{2}}\Big(v_{u}+H_{u}^0+iP_{u}^0\Big)\end{array}\right),
~~~~~~
H_{d}=\left(\begin{array}{c}{1\over\sqrt{2}}\Big(v_{d}+H_{d}^0+iP_{d}^0\Big)\\H_{d}^-\end{array}\right).
\end{eqnarray}

The three Higgs singlets are represented by:
\begin{eqnarray}
&&\chi_1={1\over\sqrt{2}}\Big(v_{\eta}+\phi_{1}^0+iP_{1}^0\Big),~~~~~~~~~~~~~~~
\chi_2={1\over\sqrt{2}}\Big(v_{\bar{\eta}}+\phi_{2}^0+iP_{2}^0\Big),\nonumber\\&&
\hspace{3.0cm}S={1\over\sqrt{2}}\Big(v_{S}+\phi_{S}^0+iP_{S}^0\Big).
\end{eqnarray}

The VEVs of the Higgs superfields $H_u$, $H_d$, $\chi_1$, $\chi_2$ and $S$ are denoted by $v_u,~v_d,~v_\eta$,~ $v_{\bar\eta}$ and $v_S$ respectively. Two angles are defined as $\tan\beta=v_u/v_d$ and $\tan\beta_\eta=v_{\bar{\eta}}/v_{\eta}$.

The soft SUSY breaking terms of N-B-LSSM are:
\begin{eqnarray}
&&\mathcal{L}_{soft}=\mathcal{L}_{soft}^{MSSM}-\frac{T_\kappa}{3}S^3+\epsilon_{ij}T_{\lambda}SH_d^iH_u^j+T_{2}S\chi_1\chi_2\nonumber\\&&
-T_{\chi,ik}\chi_1\tilde{\nu}_{R,i}^{*}\tilde{\nu}_{R,k}^{*}
+\epsilon_{ij}T_{\nu,ij}H_u^i\tilde{\nu}_{R,i}^{*}\tilde{e}_{L,j}-m_{\eta}^2|\chi_1|^2-m_{\bar{\eta}}^2|\chi_2|^2\nonumber\\&&-m_S^2|S|^2-m_{\nu,ij}^2\tilde{\nu}_{R,i}^{*}\tilde{\nu}_{R,j}
-\frac{1}{2}(2M_{BB^\prime}\lambda_{\tilde{B}}\tilde{B^\prime}+M_{BL}\tilde{B^\prime}^2)+h.c~~.\label{L}
\end{eqnarray}

The particle contents and charge assignments for N-B-LSSM are shown in the Table \ref {I}.
\begin{table}[h]
\caption{ The superfields in N-B-LSSM}
\begin{tabular}{|c|c|c|c|c|}
\hline
Superfields & $SU(3)_C$ & $SU(2)_L$ & $U(1)_Y$ & $U(1)_{B-L}$ \\
\hline
$\hat{q}$ & 3 & 2 & 1/6 & 1/6  \\
\hline
$\hat{l}$ & 1 & 2 & -1/2 & -1/2  \\
\hline
$\hat{H}_d$ & 1 & 2 & -1/2 & 0 \\
\hline
$\hat{H}_u$ & 1 & 2 & 1/2 & 0 \\
\hline
$\hat{d}$ & $\bar{3}$ & 1 & 1/3 & -1/6  \\
\hline
$\hat{u}$ & $\bar{3}$ & 1 & -2/3 & -1/6 \\
\hline
$\hat{e}$ & 1 & 1 & 1 & 1/2 \\
\hline
$\hat{\nu}$ & 1 & 1 & 0 & 1/2 \\
\hline
$\hat{\chi}_1$ & 1 & 1 & 0 & -1 \\
\hline
$\hat{\chi}_2$ & 1 & 1 & 0 & 1\\
\hline
$\hat{S}$ & 1 & 1 & 0 & 0 \\
\hline
\end{tabular}
\label{I}
\end{table}

In the theory with two Abelian groups $U(1)_Y$ and $U(1)_{B-L}$, a new effect called gauge kinetic mixing occurs. Even if the initial value of this mixing term is zero at $M_{GUT}$, non-zero values can still be generated through the evolution of RGEs.

The covariant derivatives of N-B-LSSM can be written as \cite{UMSSM5,B-L1,B-L2,gaugemass}:
{\begin{eqnarray}
&&D_\mu=\partial_\mu-i\left(\begin{array}{cc}Y^Y,&Y^{B-L}\end{array}\right)
\left(\begin{array}{cc}g_{Y},&g{'}_{{YB}}\\g{'}_{{BY}},&g{'}_{{B-L}}\end{array}\right)
\left(\begin{array}{c}B_{\mu}^{\prime Y} \\ B_{\mu}^{\prime BL}\end{array}\right)\;.
\end{eqnarray}}

$B_\mu^{\prime Y}$ and $B_\mu^{\prime BL}$ denote the gauge fields of $U(1)_Y$ and  $U(1)_{B-L}$ respectively. Under the condition that the two Abelian gauge groups are not broken, we can do a change of basis using the rotation matrix R satisfying the orthogonality condition $R^T R=1$ \cite{UMSSM5,B-L2,gaugemass}.
\begin{eqnarray}
&&\left(\begin{array}{cc}g_{Y},&g{'}_{{YB}}\\g{'}_{{BY}},&g{'}_{{B-L}}\end{array}\right)
R^T=\left(\begin{array}{cc}g_{1},&g_{{YB}}\\0,&g_{{B}}\end{array}\right)~~~~\text{and}~~~~~
R\left(\begin{array}{c}B_{\mu}^{\prime Y} \\ B_{\mu}^{\prime BL}\end{array}\right)
=\left(\begin{array}{c}B_{\mu}^{Y} \\ B_{\mu}^{BL}\end{array}\right)\;.
\end{eqnarray}

At the tree level, three neutral gauge bosons $B^{Y}_\mu,~B^{{BL}}_\mu$ and $V^3_\mu$ undergo mixing, with their mass matrix expressed in the basis $(B^{Y}_\mu, B^{{BL}}_\mu, V^3_\mu)$:
\begin{eqnarray}
&&\left(\begin{array}{*{20}{c}}
\frac{1}{8}g_{1}^2 v^2 &~~~ -\frac{1}{8}g_{1}g_{2} v^2 & ~~~\frac{1}{8}g_{1}(g_{YB}+g_B) v^2 \\
-\frac{1}{8}g_{1}g_{2} v^2 &~~~ \frac{1}{8}g_{2}^2 v^2 & ~~~~-\frac{1}{8}g_{2}(g_{YB}+g_B) v^2\\
\frac{1}{8}g_{1}(g_{YB}+g_B) v^2 &~~~ -\frac{1}{8}g_{2}(g_{YB}+g_B) v^2 &~~~~ \frac{1}{8}(g_{YB}+g_B)^2 v^2+\frac{1}{8}g_{{B}}^2 \xi^2
\end{array}\right),\label{matrix}
\end{eqnarray}
with $v^2=v_u^2+v_d^2$ and $\xi^2=v_\eta^2+v_{\bar{\eta}}^2$.

The mass eigenvalues of the matrix in Eq. (\ref{matrix}) are determined through two mixing angles: the Weinberg angle $\theta_{W}$ and a newly introduced angle $\theta_{W}'$. The latter is defined as follows:
\begin{eqnarray}
\sin^2\theta_{W}'\!=\!\frac{1}{2}\!-\!\frac{[(g_{{YB}}+g_{B})^2-g_{1}^2-g_{2}^2]v^2+
4g_{B}^2\xi^2}{2\sqrt{[(g_{{YB}}+g_{B})^2+g_{1}^2+g_{2}^2]^2v^4\!+\!8g_{B}^2[(g_{{YB}}+g_{B})^2\!-\!g_{1}^2\!-\!g_{2}^2]v^2\xi^2\!+\!16g_{B}^4\xi^4}}.
\end{eqnarray}

The new mixing angle appears in the couplings involving $Z$ and $Z^{\prime}$. The exact eigenvalues of Eq. (\ref{matrix}) are deduced:
\begin{eqnarray}
&&m_\gamma^2=0,\nonumber\\
&&m_{Z,{Z^{'}}}^2=\frac{1}{8}\Big([g_{1}^2+g_2^2+(g_{{YB}}+g_{B})^2]v^2+4g_{B}^2\xi^2 \nonumber\\
&&\hspace{1.1cm}\mp\sqrt{[g_{1}^2+g_{2}^2+(g_{{YB}}+g_{B})^2]^2v^4\!+\!8[(g_{{YB}}+g_{B})^2\!-\!g_{1}^2\!-\!
g_{2}^2]g_{B}^2v^2\xi^2\!+\!16g_{B}^4\xi^4}\Big).
\end{eqnarray}

In the calculation, the mass squared matrices of the neutralino, chargino, slepton, CP-even sneutrino, CP-odd sneutrino, up squark and down squark are required. These mass matrices can be found in Refs. \cite{han,SRZ}.

Here, we show some needed couplings in this model. The Z bosons interact with sneutrinos, whose explicit form reads as:
\begin{eqnarray}
&&\mathcal{L}_{Z\tilde{\nu}^I\tilde{\nu}^R}=\frac{1}{2}\tilde{\nu}^I_i
\Big[\Big(g_1\cos\theta_W^\prime\sin\theta_W+g_2\cos\theta_W\cos\theta_W^\prime
-(g_{YB}+g_B)\sin\theta_W^\prime\Big)\sum_{a=1}^3Z_{i,a}^{I,*}Z_{j,a}^{R,*}\nonumber\\&&\hspace{1.5cm}
-g_B\sin\theta_W^\prime\sum_{a=1}^3Z_{i,3+a}^{I,*}Z_{j,3+a}^{R,*}\Big](p^{\nu^I_{i}}_\mu-p^{\nu^R_{j}}_\mu)\tilde{\nu}^R_jZ_{\mu}.
\end{eqnarray}

We also deduce the vertex of $Z-{\chi}_i^{0}-{\chi}_j^{0}$:
\begin{eqnarray}
&&\mathcal{L}_{Z {\chi}^{0} {\chi}^{0}}= \frac{i}{2}{\chi}^{0}_i\Big\{ \Big[(g_{YB}\sin\theta_{W}'-g_1 \cos\theta_{W}' \sin\theta_{W} -g_{2} \cos\theta_{W} \cos\theta_{W}')(N_{j3}^*N_{i3}-N_{j4}^*N_{i4})\nonumber\\&&\hspace{1.4cm}+2g_{B}\sin\theta_{W}'(N_{j6}^*N_{i6}-N_{j7}^*N_{i7})\Big] {\gamma}_{\mu}P_L\nonumber\\&&\hspace{1.4cm}+  \Big[(g_1 \cos\theta_{W}' \sin\theta_{W} +g_{2} \cos\theta_{W} \cos\theta_{W}'-g_{YB}\sin\theta_{W}')(N_{i3}^*N_{j3}-N_{i4}^*N_{j4})\nonumber\\&&\hspace{1.4cm}-2g_{B}\sin\theta_{W}'(N_{i6}^*N_{j6}-N_{i7}^*N_{j7})\Big] {\gamma}_{\mu}P_R\Big\}{\chi}^{0}_j Z_{\mu}.
\end{eqnarray}

The vertices of $Z-\bar{d}_i-d_j$ and $Z-\bar{u}_i-u_j$ are:
\begin{eqnarray}
&&\mathcal{L}_{Z d\bar{d}}=\frac{i}{6}\bar{d}_i \Big\{\Big[3g_2\cos\theta_{W}\cos\theta_{W}'+g_1\cos\theta_{W}' \sin\theta_{W}-(g_{YB}+g_B) \sin\theta_{W}'\Big]{\gamma}_{\mu}P_L\nonumber\\
&&\hspace{1.4cm}+\Big[(2g_{YB}-g_B)\sin\theta_{W}'-2g_1\cos\theta_{W}' \sin\theta_{W}\Big]{\gamma}_{\mu}P_R\Big\}d_jZ_{\mu},
\end{eqnarray}
\begin{eqnarray}
&&\mathcal{L}_{Z u\bar{u}}=\frac{i}{6}\bar{u}_i \Big\{\Big[g_1\cos\theta_{W}'\sin\theta_{W}-3g_2\cos\theta_{W} \cos\theta_{W}'-(g_{YB}+g_B) \sin\theta_{W}'\Big]{\gamma}_{\mu}P_L\nonumber\\
&&\hspace{1.4cm}+\Big[4g_1\cos\theta_{W}'\sin\theta_{W}-(4g_{YB}+g_B)\sin\theta_{W}'\Big]{\gamma}_{\mu}P_R\Big\}u_jZ_{\mu}.
\end{eqnarray}

To save space in the text, the remaining vertices can be found in Ref. \cite{SRZ}.

\section{Analytical formula}

In this section, we systematically study the amplitudes and branching ratios of the LFV processes $\tau \rightarrow e \pi^+\pi^-$, $\tau\rightarrow e\pi^+K^-$, $\tau \rightarrow e K^+ K^-$ in the N-B-LSSM. To ensure a comprehensive analysis, we construct all the relevant Feynman diagrams, including penguin-type, self-energy-type and box-type diagrams. Next, we give the effective amplitudes of the processes at the quark level.

\subsection{the penguin-type diagrams}

When the external leptons are all on shell, and the required Wilson coefficients are extracted, the contribution from the $\gamma$-penguin-type diagram in Fig.\ref{penguin}(a) can be written as:
\begin{eqnarray}
&&\mathcal{M}_{\gamma-p}^{(a)} =\frac{-Q_q e^2}{k^2}\sum_{F=\chi^0, \chi^\pm} \sum_{S=\tilde{e},\tilde{\nu}}\{\frac{1}{2} I_1(x_F,x_S) H_R^{S F \bar{e}} H_L^{S^* \tau \bar{F}} \nonumber\\&&+[I_2(x_F,x_S)-I_3(x_F,x_S)][m_F(m_e H_L^{S F \bar{e}} H_L^{S^* \tau \bar{F}}
+m_\tau H_R^{S F \bar{e}} H_R^{S^* \tau \bar{F}} )]\nonumber\\&&
+[I_2(x_F,x_S)-I_4(x_F,x_S)][m_e m_\tau H_L^{S F \bar{e}} H_R^{S^* \tau \bar{F}} +(m^2_\tau+m^2_e)H_R^{S F \bar{e}} H_L^{S^* \tau \bar{F}} ]\}\nonumber\\&&
\times (\bar{e} \gamma^\mu P_L \tau)(\bar{q} \gamma_\mu P_L q+\bar{q} \gamma_\mu P_R q)+(L \leftrightarrow R).
\end{eqnarray}
where $P_L=\frac{1-\gamma_5}{2}$, $P_R=\frac{1+\gamma_5}{2}$, $Q_u=\frac{2}{3}$, $Q_d=-\frac{1}{3}$, respectively. Additionally, $x_i=\frac{m_i^2}{\Lambda^2}$ and $m_i$ denote the mass of the corresponding particle, $\Lambda$ represents the energy scale of NP and $k$ expresses the characteristic energy scale of QCD. $H_{L,R}^{S F \bar{e}}$ and $H_{L,R}^{S^* \tau \bar{F}}$ are the corresponding couplings of the left(right)-hand parts in the Lagrangian. The concrete expressions for form factors $I_i$ ($i$=1,...,4) are collected here:
\begin{figure}
\setlength{\unitlength}{5.0mm}
\centering
\includegraphics[width=5.0in]{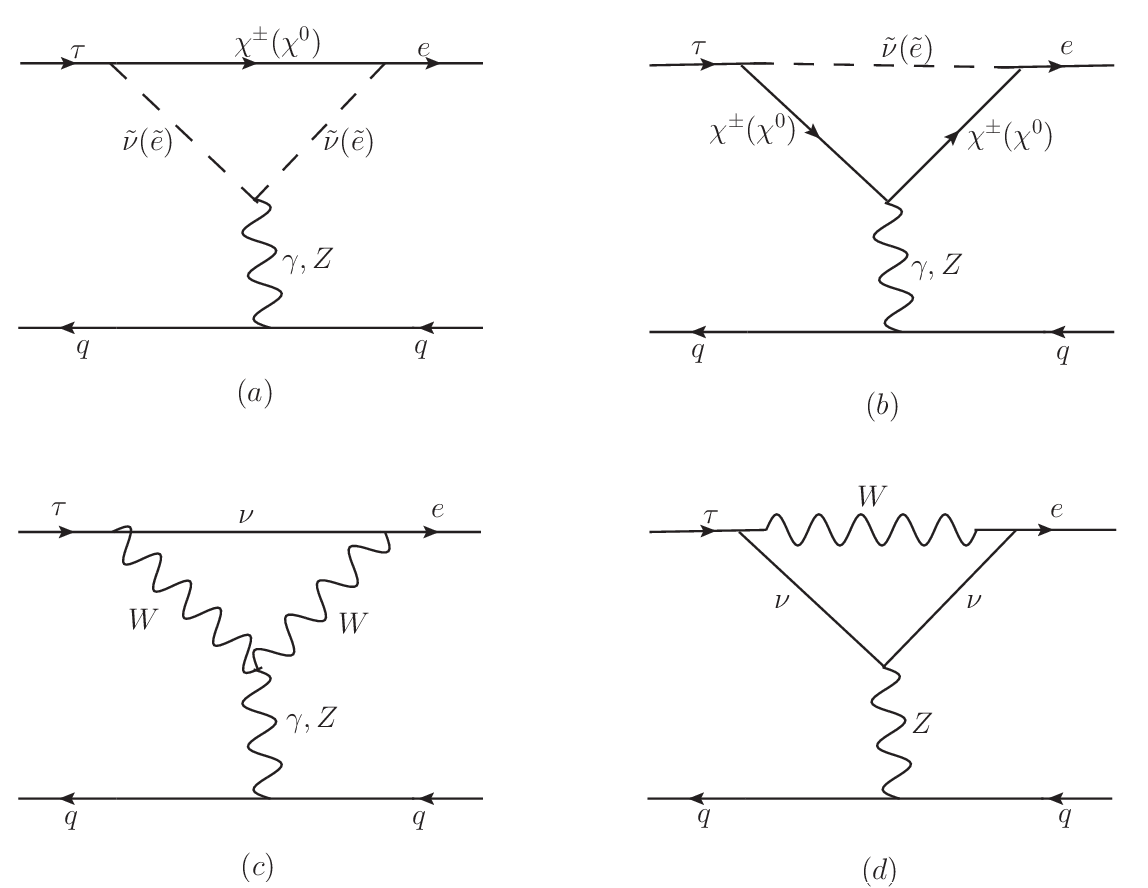}
\caption{The penguin-type diagrams for LFV processes $\tau \rightarrow e \pi^+\pi^-$, $\tau\rightarrow e\pi^+K^-$ and $\tau \rightarrow e K^+ K^-$ in the N-B-LSSM.}\label{penguin}
\end{figure}
\begin{eqnarray}
&&I_1(x_1,x_2)=\frac{1}{16\pi^2}\Big[\frac{x_1}{x_1-x_2}+\frac{(2x_1-x_2)x_2 \ln x_2-x_1^2 \ln x_1}{(x_1-x_2)^2}\Big],\nonumber\\&&
I_2(x_1,x_2)=\frac{1}{96\Lambda^2\pi^2}\Big[\frac{11x_1^2-7x_1x_2+2x_2^2}{(x_1-x_2)^3}+\frac{6x_1^3(\ln x_2-\ln x_1)}{(x_1-x_2)^4}\Big],\nonumber\\&&
I_3(x_1,x_2)=\frac{1}{16\Lambda^2\pi^2}\Big[\frac{1}{x_1-x_2}+\frac{x_1(\ln x_2-\ln x_1)}{(x_1-x_2)^2}\Big],\nonumber\\&&
I_4(x_1,x_2)=\frac{1}{32\Lambda^2\pi^2}\Big[\frac{3x_1-x_2}{(x_1-x_2)^2}+\frac{2x_1^2(\ln x_2-\ln x_1)}{(x_1-x_2)^3}\Big].
\end{eqnarray}

Similarly, the effective amplitude from the $\gamma$-penguin-type diagram drawn in Fig.\ref{penguin}(b) is in the following:
\begin{eqnarray}
&&\mathcal{M}_{\gamma-p}^{(b)} = \frac{Q_q e^2}{k^2}\sum_{F=\chi^0, \chi^\pm} \sum_{S=\tilde{e},\tilde{\nu}}\{[\frac{1}{2}I_1(x_F,x_S)-m^2_F I_3(x_F,x_S)] H_R^{S F \bar{e}} H_L^{S^* \tau \bar{F}}\nonumber\\&&
+[I_4(x_F,x_S)-I_3(x_F,x_S)][m_F(m_e H_L^{S F \bar{e}}  H_L^{S^* \tau \bar{F}} +m_\tau H_R^{S F \bar{e}}  H_R^{S^* \tau \bar{F}} )]\nonumber\\&&
+[2I_4(x_F,x_S)-I_2(x_F,x_S)-I_3(x_F,x_S)][m_e m_\tau H_L^{S F \bar{e}}  H_R^{S^* \tau \bar{F}} \nonumber\\&&+(m^2_\tau+m^2_e)
H_R^{S F \bar{e}}  H_L^{S^* \tau \bar{F}} ]\}
\times (\bar{e} \gamma^\mu P_L \tau)(\bar{q} \gamma_\mu P_L q+\bar{q} \gamma_\mu P_R q)+(L \leftrightarrow R).
\end{eqnarray}

The contributions from $Z$-penguin-type diagrams are derived in the same way as $\gamma$-penguin-type diagrams:
\begin{eqnarray}
&&\mathcal{M}_{Z-p}^{(a)} = \sum_{F=\chi^0, \chi^\pm} \sum_{S=\tilde{e},\tilde{\nu}}\{\frac{1}{2 m^2_Z}I_5(x_F,x_{S_1},x_{S_2}) [H_R^{S_2 F \bar{e}}H_L^{S^* \tau \bar{F}}H^{Z S_1 S^*_1}H_L^{\bar{q} Z q}\nonumber\\&&~~~~~~~~
\times(\bar{e} \gamma^\mu P_L \tau)(\bar{q} \gamma_\mu P_L q)+H_R^{S_2 F \bar{e}}H_L^{S^* \tau \bar{F}}H^{Z S_1 S^*_1}H_R^{\bar{q} Z q}\nonumber\\&&~~~~~~~~
\times(\bar{e} \gamma^\mu P_L \tau)(\bar{q} \gamma_\mu P_R q)]+(L \leftrightarrow R)\}.
\end{eqnarray}
\begin{eqnarray}
&&\mathcal{M}_{Z-p}^{(b)} = \sum_{F=\chi^0, \chi^\pm} \sum_{S=\tilde{e},\tilde{\nu}}\{ -\frac{1}{2 m^2_Z}I_5(x_S,x_{F_2},x_{F_1}) [H_R^{S F_2 \bar{e}}H_R^{Z F_1 \bar{F_2}}H_L^{S^* \tau \bar{F_1}}H_L^{\bar{q} Z q}\nonumber\\&&~~~~~~~~~\times(\bar{e} \gamma^\mu P_L \tau)(\bar{q} \gamma_\mu P_L q)+H_R^{S F_2 \bar{e}}H_R^{Z F_1 \bar{F_2}}H_L^{S^* \tau \bar{F_1}}H_R^{\bar{q} Z q}\times(\bar{e} \gamma^\mu P_L \tau)(\bar{q} \gamma_\mu P_R q)]\nonumber\\&&~~~~~~~~~
+\frac{m_{F_1}m_{F_2}}{m^2_Z}I_6(x_S,x_{F_2},x_{F_1})[H_R^{S F_2 \bar{e}}H_L^{Z F_1 \bar{F_2}}H_L^{S^* \tau \bar{F_1}}H_L^{\bar{q} Z q}
\times(\bar{e} \gamma^\mu P_L \tau)(\bar{q} \gamma_\mu P_L q)\nonumber\\&&~~~~~~~~~+H_R^{S F_2 \bar{e}}H_L^{Z F_1 \bar{F_2}}H_L^{S^* \tau \bar{F_1}}H_R^{\bar{q} Z q}
\times(\bar{e} \gamma^\mu P_L \tau)(\bar{q} \gamma_\mu P_R q)]
+(L \leftrightarrow R)\}.
\end{eqnarray}

The functions $I_5(x_1,x_2,x_3)$ and $I_6(x_1,x_2,x_3)$ are:
\begin{eqnarray}
&&I_5(x_1,x_2,x_3)\nonumber\\&&~~~~~~~~~=\frac{1}{16\pi^2}\Big[1-\frac{x_1^2\ln x_1}{(x_1-x_2)(x_1-x_3)}+\frac{x_2^2\ln x_2}{(x_1-x_2)(x_2-x_3)}-\frac{x_3^2\ln x_3}{(x_1-x_3)(x_2-x_3)}\Big],\nonumber\\&&
I_6(x_1,x_2,x_3)\nonumber\\&&~~~~~~~~~=\frac{1}{16\Lambda^2\pi^2}\Big[\frac{x_1\ln x_1}{(x_1-x_2)(x_1-x_3)}-\frac{x_2\ln x_2}{(x_1-x_2)(x_2-x_3)}+\frac{x_3\ln x_3}{(x_1-x_3)(x_2-x_3)}\Big].
\end{eqnarray}

After detailed analysis, we conclude that the contribution from the $W-\nu$ diagrams can be reasonably neglected under the current model and energy scale, primarily for the following reason: In the N-B-LSSM, LFV processes of the $W-\nu$ diagrams mainly originate from neutrinos Yukawa couplings $Y_\nu$. In the rotation matrix $Z^\nu$ introduced by the diagonalisation of the neutrino mass matrix, the magnitude of the off-diagonal element $Z^{\nu}_{i\neq j}$ can be approximately estimated as $Z^\nu_{i\neq j}= \frac{v_u}{2 v_\eta}\cdot\frac{Y_\nu ^{T}}{Y_X }$. Because $Y^{ij}_{\nu} (i\neq j)$ itself is extremely small (typically $\lesssim 10^{-6}$), and $v_u \ll v_\eta$, the off-diagonal element is usually at the order of $10^{-9}$ or even smaller. The $W-\nu$ diagram brings in a $Z^{\nu}_{i\neq j}$ at each vertex, and its overall contribution is $Z^\nu_{ij}Z^\nu_{ji}\sim 10^{-18}$. Based on the above consideration, we prioritize the calculation of other diagrams, particularly those involving contributions from NP, while omitting further computations for the $W-\nu$ diagrams. Similar to the penguin-type diagrams, the corrections generated by $W-\nu$ through the self-energy-type and box-type diagrams are not presented to save space.
\subsection{the self-energy-type diagrams}

We show the specific contribution form of the self-energy-type diagrams in Fig.\ref{self-energy}. The $\gamma$-self-energy-type diagrams give the terms:
\begin{figure}
\setlength{\unitlength}{5.0mm}
\centering
\includegraphics[width=5.0in]{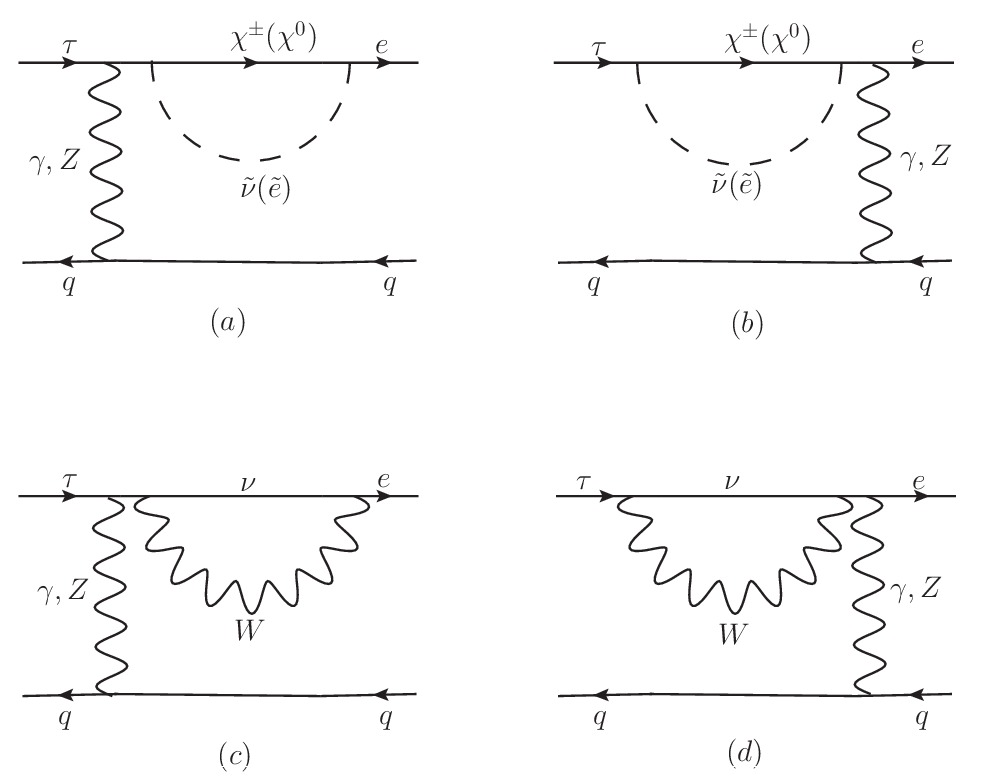}
\caption{The self-energy-type diagrams for LFV processes $\tau \rightarrow e \pi^+\pi^-$, $\tau\rightarrow e\pi^+K^-$ and $\tau \rightarrow e K^+ K^-$ in the N-B-LSSM.}\label{self-energy}
\end{figure}
\begin{eqnarray}
&&\mathcal{M}_{\gamma-S}^{(a)} =\frac{Q_q e^2}{k^2(m^2_e-m^2_\tau)}\sum_{F=\chi^0, \chi^\pm} \sum_{S=\tilde{e},\tilde{\nu}}\Big\{\frac{1}{2}I_1(x_F,x_S)[m_e(m_e H_R^{S^* F \bar{e}} H_L^{S \tau \bar{F}}
\nonumber\\&&~~~~~~~~~+m_\tau H_L^{S^* F \bar{e}} H_R^{S \tau \bar{F}} )]+\{m^2_e[I_3(x_F,x_S)-I_4(x_F,x_S)]-I_7(x_F,x_S)\} \nonumber\\&&~~~~~~~~~\times[m_F(m_e  H_L^{S^* F \bar{e}} H_L^{S \tau \bar{F}}+m_\tau H_R^{S^* F \bar{e}} H_R^{S \tau \bar{F}})]\Big\}\times(\bar{e} \gamma^\mu P_L \tau)(\bar{q} \gamma_\mu P_L q+\bar{q} \gamma_\mu P_R q)\nonumber\\&&~~~~~~~~~+(L \leftrightarrow R).
\end{eqnarray}
\begin{eqnarray}
&&\mathcal{M}_{\gamma-S}^{(b)} =\frac{Q_q e^2}{k^2(m^2_\tau-m^2_e)}\sum_{F=\chi^0, \chi^\pm} \sum_{S=\tilde{e},\tilde{\nu}}\Big\{\frac{1}{2}I_1(x_F,x_S)[m_\tau(m_\tau H_R^{S^* F \bar{e}} H_L^{S \tau \bar{F}}
\nonumber\\&&~~~~~~~~~+m_e H_L^{S^* F \bar{e}} H_R^{S \tau \bar{F}} )]+\{m^2_\tau[I_3(x_F,x_S)-I_4(x_F,x_S)]-I_7(x_F,x_S)\} \nonumber\\&&~~~~~~~~~\times[m_F(m_e  H_L^{S^* F \bar{e}} H_L^{S \tau \bar{F}}+m_\tau H_R^{S^* F \bar{e}} H_R^{S \tau \bar{F}})]\Big\}\times(\bar{e} \gamma^\mu P_L \tau)(\bar{q} \gamma_\mu P_L q+\bar{q} \gamma_\mu P_R q)\nonumber\\&&~~~~~~~~~+(L \leftrightarrow R).
\end{eqnarray}
with
\begin{eqnarray}
&&I_7(x_1,x_2)=\frac{1}{16\pi^2}\Big[1+\frac{x_1\ln x_1}{x_2-x_1}+\frac{x_2 \ln x_2}{x_1-x_2}\Big].
\end{eqnarray}

Furthermore, the effective amplitudes from the Z-self-energy-type diagrams drawn in Fig.\ref{self-energy} can be written as:
\begin{eqnarray}
&&\mathcal{M}_{Z-S}^{(a)} =\frac{1}{(k^2-m^2_Z)(m^2_e-m^2_\tau)}\sum_{F=\chi^0, \chi^\pm} \sum_{S=\tilde{e},\tilde{\nu}}\Big\{-\frac{1}{2}I_1(x_F,x_S)[m_e(m_e H_R^{S^* F \bar{e}} H_L^{S \tau \bar{F}}H_L^{Z \tau \bar{e}}H_L^{\bar{q} Z q}
\nonumber\\&&~~~~~~~~~+m_\tau H_L^{S^* F \bar{e}} H_R^{S \tau \bar{F}}H_L^{Z \tau \bar{e}}H_L^{\bar{q} Z q} )]+\{m^2_e[I_4(x_F,x_S)-I_3(x_F,x_S)]+I_7(x_F,x_S)\} \nonumber\\&&~~~~~~~~~\times[m_F(m_e  H_L^{S^* F \bar{e}} H_L^{S \tau \bar{F}}H_L^{Z \tau \bar{e}}H_L^{\bar{q} Z q}+m_\tau H_R^{S^* F \bar{e}} H_R^{S \tau \bar{F}}H_L^{Z \tau \bar{e}}H_L^{\bar{q} Z q})]\Big\}\nonumber\\&&~~~~~~~~~\times(\bar{e} \gamma^\mu P_L \tau)(\bar{q} \gamma_\mu P_L q)+\Big\{-\frac{1}{2}I_1(x_F,x_S)[m_e(m_e H_R^{S^* F \bar{e}} H_L^{S \tau \bar{F}}H_L^{Z \tau \bar{e}}H_R^{\bar{q} Z q}
\nonumber\\&&~~~~~~~~~+m_\tau H_L^{S^* F \bar{e}} H_R^{S \tau \bar{F}}H_L^{Z \tau \bar{e}}H_R^{\bar{q} Z q} )]+\{m^2_e[I_4(x_F,x_S)-I_3(x_F,x_S)]+I_7(x_F,x_S)\} \nonumber\\&&~~~~~~~~~\times[m_F(m_e  H_L^{S^* F \bar{e}} H_L^{S \tau \bar{F}}H_L^{Z \tau \bar{e}}H_R^{\bar{q} Z q}+m_\tau H_R^{S^* F \bar{e}} H_R^{S \tau \bar{F}}H_L^{Z \tau \bar{e}}H_R^{\bar{q} Z q})]\Big\}\nonumber\\&&~~~~~~~~~\times(\bar{e} \gamma^\mu P_L \tau)(\bar{q} \gamma_\mu P_R q)+(L \leftrightarrow R).
\end{eqnarray}
\begin{eqnarray}
&&\mathcal{M}_{Z-S}^{(b)} =\frac{1}{(k^2-m^2_Z)(m^2_\tau-m^2_e)}\sum_{F=\chi^0, \chi^\pm} \sum_{S=\tilde{e},\tilde{\nu}}\Big\{-\frac{1}{2}I_1(x_F,x_S)[m_\tau(m_\tau H_L^{Z \tau \bar{e}}H_R^{S^* F \bar{e}} H_L^{S \tau \bar{F}}H_L^{\bar{q} Z q}
\nonumber\\&&~~~~~~~~~+m_e H_L^{Z \tau \bar{e}}H_L^{S^* F \bar{e}} H_R^{S \tau \bar{F}}H_L^{\bar{q} Z q} )]+\{m^2_\tau[I_4(x_F,x_S)-I_3(x_F,x_S)]+I_7(x_F,x_S)\} \nonumber\\&&~~~~~~~~~\times[m_F(m_\tau  H_L^{Z \tau \bar{e}}H_R^{S^* F \bar{e}} H_R^{S \tau \bar{F}}H_L^{\bar{q} Z q}+m_e H_L^{Z \tau \bar{e}}H_L^{S^* F \bar{e}} H_L^{S \tau \bar{F}}H_L^{\bar{q} Z q})]\Big\}\nonumber\\&&~~~~~~~~~\times(\bar{e} \gamma^\mu P_L \tau)(\bar{q} \gamma_\mu P_L q)+\Big\{-\frac{1}{2}I_1(x_F,x_S)[m_\tau(m_\tau H_L^{Z \tau \bar{e}}H_R^{S^* F \bar{e}} H_L^{S \tau \bar{F}}H_R^{\bar{q} Z q}
\nonumber\\&&~~~~~~~~~+m_e H_L^{Z \tau \bar{e}}H_L^{S^* F \bar{e}} H_R^{S \tau \bar{F}}H_R^{\bar{q} Z q} )]+\{m^2_\tau[I_4(x_F,x_S)-I_3(x_F,x_S)]+I_7(x_F,x_S)\} \nonumber\\&&~~~~~~~~~\times[m_F(m_\tau  H_L^{Z \tau \bar{e}}H_R^{S^* F \bar{e}} H_R^{S \tau \bar{F}}H_R^{\bar{q} Z q}+m_e H_L^{Z \tau \bar{e}}H_L^{S^* F \bar{e}} H_L^{S \tau \bar{F}}H_R^{\bar{q} Z q})]\Big\}\nonumber\\&&~~~~~~~~~\times(\bar{e} \gamma^\mu P_L \tau)(\bar{q} \gamma_\mu P_R q)+(L \leftrightarrow R).
\end{eqnarray}

\subsection{the box-type diagrams}

The box-type diagrams contributing to LFV processes $\tau \rightarrow e \pi^+\pi^-$, $\tau\rightarrow e\pi^+K^-$ and $\tau \rightarrow e K^+ K^-$ in the N-B-LSSM are shown in Fig.\ref{box}. Fierz rearrangement is carried out in the calculation processes. Fig.\ref{box}(a)(b) represent the contributions from neutral fermions $\chi^0$, charged scalars $\tilde{e}$ and squark $\tilde{q}$ ($\tilde{q}$=$\tilde{u},~\tilde{d}$). We analyze the effective amplitudes $\mathcal{M}_{(n)}^{(a)}$ and $\mathcal{M}_{(n)}^{(b)}$ originate from those box diagrams with virtual neutral fermion contributions in a concrete form:
\begin{figure}
\setlength{\unitlength}{5.0mm}
\centering
\includegraphics[width=5.0in]{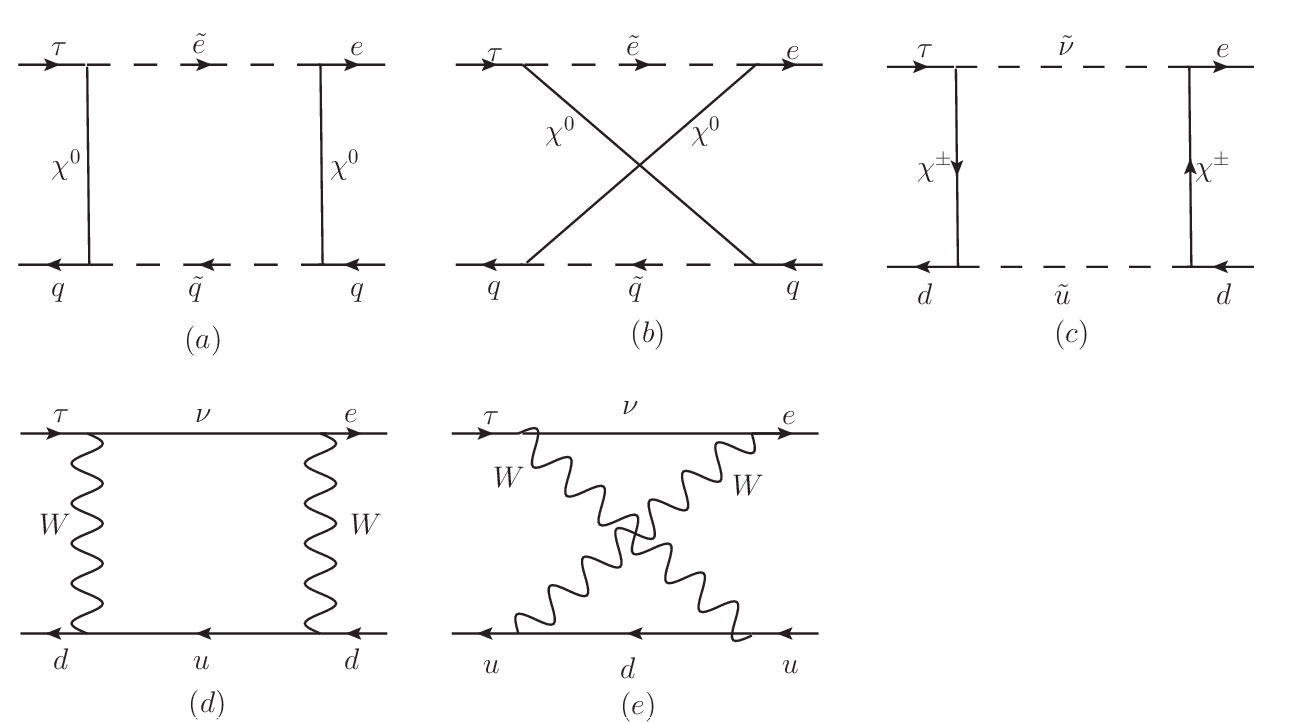}
\caption{The box-type diagrams for LFV processes $\tau \rightarrow e \pi^+\pi^-$, $\tau\rightarrow e\pi^+K^-$ and $\tau \rightarrow e K^+ K^-$ in the N-B-LSSM.}\label{box}
\end{figure}
\begin{eqnarray}
&&\mathcal{M}_{(n)}^{(a)}=\sum_{F_1,F_2=\chi^0,\chi^0} \sum_{S_1,S_2=\tilde{e},\tilde{q}}\Big\{\frac{1}{2}I_8(x_{F_1},x_{F_2},x_{S_1},x_{S_2})\Big[\frac{1}{2}H_L^{S^*_1\tau\bar{F_1}}H_R^{S_1 F_2\bar{e}}H_R^{S_2 F_1 \bar{q}}H_L^{S^*_2 q\bar{F_2}}\nonumber\\&&\times(\bar{e} \gamma^\mu P_L \tau)(\bar{q} \gamma_\mu P_L q)
-H_R^{S^*_1\tau\bar{F_1}}H_R^{S_1 F_2\bar{e}}H_L^{S_2 F_1 \bar{q}}H_L^{S^*_2 q\bar{F_2}}\times(\bar{e} P_R \tau)(\bar{q} P_L q)\Big]\nonumber\\&&+\frac{1}{8}I_9(x_{F_1},x_{F_2},x_{S_1},x_{S_2})
\Big[m_{F_1}m_{F_2}H_L^{S^*_1\tau\bar{F_1}}H_L^{S_1 F_2\bar{e}}H_L^{S_2 F_1 \bar{q}}H_L^{S^*_2 q\bar{F_2}}\nonumber\\&&\times[(\bar{e} P_R \tau)(\bar{q} P_R q)
-(\bar{e} P_R \tau)(\bar{q} P_L q)-(\bar{e} P_L \tau)(\bar{q} P_R q)-3(\bar{e} P_L \tau)(\bar{q} P_L q)\nonumber\\&&-(\bar{e}\sigma_{\mu\nu}P_L \tau)(\bar{q} \sigma^{\mu\nu}P_L q)
-(\bar{e}\sigma_{\mu\nu}P_R \tau)(\bar{q} \sigma^{\mu\nu}P_L q)]+H_R^{S^*_1\tau\bar{F_1}}H_L^{S_1 F_2\bar{e}}H_R^{S_2 F_1 \bar{q}}H_L^{S^*_2 q\bar{F_2}}\nonumber\\&&\times[-4(\bar{e} \gamma^\mu P_R \tau)(\bar{q} \gamma_\mu P_L q)
+(\bar{e} P_R \tau)(\bar{q} P_R q)-(\bar{e} P_R \tau)(\bar{q} P_L q)\nonumber\\&&-(\bar{e} P_L \tau)(\bar{q} P_R q)+(\bar{e} P_L \tau)(\bar{q} P_L q)]\Big]
+(L \leftrightarrow R)\Big\}.
\end{eqnarray}
\begin{eqnarray}
&&\mathcal{M}_{(n)}^{(b)}=\sum_{F_1,F_2=\chi^0,\chi^0}
\sum_{S_1,S_2=\tilde{e},\tilde{q}}\Big\{\frac{1}{2}I_8(x_{F_1},x_{F_2},x_{S_1},x_{S_2})\Big[\frac{1}{2}H_L^{S^*_1\tau\bar{F_1}}H_R^{S_1 F_2\bar{e}}H_L^{S_2 F_2 \bar{q}}H_R^{S^*_2 q \bar{F_1 }}\nonumber\\&&\times
(\bar{e} \gamma^\mu P_L \tau)(\bar{q} \gamma_\mu P_R q)+H_L^{S^*_1\tau\bar{F_1}}H_L^{S_1 F_2\bar{e}}H_R^{S_2 F_2 \bar{q}}H_R^{S^*_2 q \bar{F_1 }}\times(\bar{e} P_L \tau)(\bar{q} P_R q)\Big]\nonumber\\&&+I_9(x_{F_1},x_{F_2},x_{S_1},x_{S_2})
\Big[\frac{m_{F_1}m_{F_2}}{2}H_L^{S^*_1\tau\bar{F_1}}H_L^{S_1 F_2\bar{e}}H_L^{S_2 F_2 \bar{q}}H_L^{S^*_2 q \bar{F_1 }}(\bar{e} P_L \tau)(\bar{q} P_L q)\nonumber\\&&+\frac{1}{8}H_L^{S^*_1\tau\bar{F_1}}H_L^{S_1 F_2\bar{e}}H_L^{S_2 F_2 \bar{q}}H_L^{S^*_2 q \bar{F_1 }}
[(\bar{e} P_L \tau)(\bar{q} P_R q)+(\bar{e} P_R \tau)(\bar{q} P_L q)-(\bar{e} P_R \tau)(\bar{q} P_R q)\nonumber\\&&-(\bar{e} P_L \tau)(\bar{q} P_L q)-(\bar{e}\sigma_{\mu\nu}P_R \tau)(\bar{q} \sigma^{\mu\nu}P_L q)
-(\bar{e}\sigma_{\mu\nu}P_L \tau)(\bar{q} \sigma^{\mu\nu}P_L q)]\nonumber\\&&+\frac{1}{8}H_R^{S^*_1\tau\bar{F_1}}H_L^{S_1 F_2\bar{e}}H_L^{S_2 F_2 \bar{q}}H_R^{S^*_2 q \bar{F_1 }}[(\bar{e} P_R \tau)(\bar{q} P_L q)+(\bar{e} P_L \tau)(\bar{q} P_R q)\nonumber\\&&
-(\bar{e} P_L \tau)(\bar{q} P_L q)-(\bar{e} P_R \tau)(\bar{q} P_R q)
-4(\bar{e} \gamma^\mu P_R \tau)(\bar{q} \gamma_\mu P_R q)]\Big]+(L \leftrightarrow R)\Big\}.
\end{eqnarray}

The concrete expressions for the functions $I_8(x_1,x_2,x_3,x_4)$ and $I_9(x_1,x_2,x_3,x_4)$ are defined as follows:
\begin{eqnarray}
&&I_8(x_1,x_2,x_3,x_4)=\frac{1}{16\Lambda^2\pi^2}\Big[\frac{x_1^2 \ln x_1}{(x_1-x_2)(x_1-x_3)(x_1-x_4)}-\frac{x_2^2 \ln x_2}{(x_1-x_2)(x_2-x_3)(x_2-x_4)}\nonumber\\&&~~~~~~~~~~~~~~~~~~~~
+\frac{x_3^2 \ln x_3}{(x_1-x_3)(x_2-x_3)(x_3-x_4)}-\frac{x_4^2 \ln x_4}{(x_1-x_4)(x_2-x_4)(x_3-x_4)}\Big].\nonumber\\&&
I_9(x_1,x_2,x_3,x_4)=\frac{1}{16\Lambda^4\pi^2}\Big[-\frac{x_1 \ln x_1}{(x_1-x_2)(x_1-x_3)(x_1-x_4)}+\frac{x_2\ln x_2}{(x_1-x_2)(x_2-x_3)(x_2-x_4)}\nonumber\\&&~~~~~~~~~~~~~~~~~~~~
-\frac{x_3 \ln x_3}{(x_1-x_3)(x_2-x_3)(x_3-x_4)}+\frac{x_4 \ln x_4}{(x_1-x_4)(x_2-x_4)(x_3-x_4)}\Big].
\end{eqnarray}

Correspondingly, Fig.\ref{box}(c) represents the contribution from charged fermions $\chi^{\pm}$, neutral scalars $\tilde{\nu}$ and squark $\tilde{u}$. The effective amplitude $\mathcal{M}_{(c)}$ from the box diagram with virtual charged fermion contribution is:
\begin{eqnarray}
&&\mathcal{M}_{(c)}=\sum_{F_1,F_2=\chi^{\pm},\chi^{\pm}} \sum_{S_1,S_2=\tilde{\nu},\tilde{u}}\Big\{\frac{1}{2}I_8(x_{F_1},x_{F_2},x_{S_1},x_{S_2})\Big[\frac{1}{2}H_L^{S_1\tau\bar{F_1}}H_R^{S_1 F_2\bar{e}}H_R^{S_2 F_1 \bar{d}}H_L^{S_2 d\bar{F_2}}\nonumber\\&&\times(\bar{e} \gamma^\mu P_L \tau)(\bar{d} \gamma_\mu P_L d)
-H_R^{S_1\tau\bar{F_1}}H_R^{S_1 F_2\bar{e}}H_L^{S_2 F_1 \bar{d}}H_L^{S_2 d\bar{F_2}}\times(\bar{e} P_R \tau)(\bar{d} P_L d)\Big]\nonumber\\&&+\frac{1}{8}I_9(x_{F_1},x_{F_2},x_{S_1},x_{S_2})
\Big[m_{F_1}m_{F_2}H_L^{S_1\tau\bar{F_1}}H_L^{S_1 F_2\bar{e}}H_L^{S_2 F_1 \bar{d}}H_L^{S_2 d\bar{F_2}}\nonumber\\&&\times[(\bar{e} P_R \tau)(\bar{d} P_R d)
-(\bar{e} P_R \tau)(\bar{d} P_L d)-(\bar{e} P_L \tau)(\bar{d} P_R d)-3(\bar{e} P_L \tau)(\bar{d} P_L d)\nonumber\\&&-(\bar{e}\sigma_{\mu\nu}P_L \tau)(\bar{d} \sigma^{\mu\nu}P_L d)
-(\bar{e}\sigma_{\mu\nu}P_R \tau)(\bar{d} \sigma^{\mu\nu}P_L d)]+H_R^{S_1\tau\bar{F_1}}H_L^{S_1 F_2\bar{e}}H_R^{S_2 F_1 \bar{d}}H_L^{S_2 d\bar{F_2}}\nonumber\\&&\times[-4(\bar{e} \gamma^\mu P_R \tau)(\bar{d} \gamma_\mu P_L d)
+(\bar{e} P_R \tau)(\bar{d} P_R d)-(\bar{e} P_R \tau)(\bar{d} P_L d)\nonumber\\&&-(\bar{e} P_L \tau)(\bar{d} P_R d)+(\bar{e} P_L \tau)(\bar{d} P_L d)]\Big]
+(L \leftrightarrow R)\Big\}.
\end{eqnarray}

\subsection{Using MIA to calculate $\tau\rightarrow e M^+M^-$}

In this work, we primarily adopt the mass eigenstate method for our calculations. This method allows for a systematic and precise treatment of particle mixing, mass spectra and complete one-loop contributions. It enables us to include all physical states and their interactions comprehensively, thus ensuring the rigor and accuracy of our results. However, due to the involvement of multiple mixing matrices, rotation matrices and mass eigenvalues, the resulting expressions are often complex, making it less straightforward to identify which parameters play the dominant role in LFV processes. This complexity can obscure the physical intuition, especially regarding the sensitivity of specific parameters.

\begin{figure}[ht]
\setlength{\unitlength}{5mm}
\centering
\includegraphics[width=5.0in]{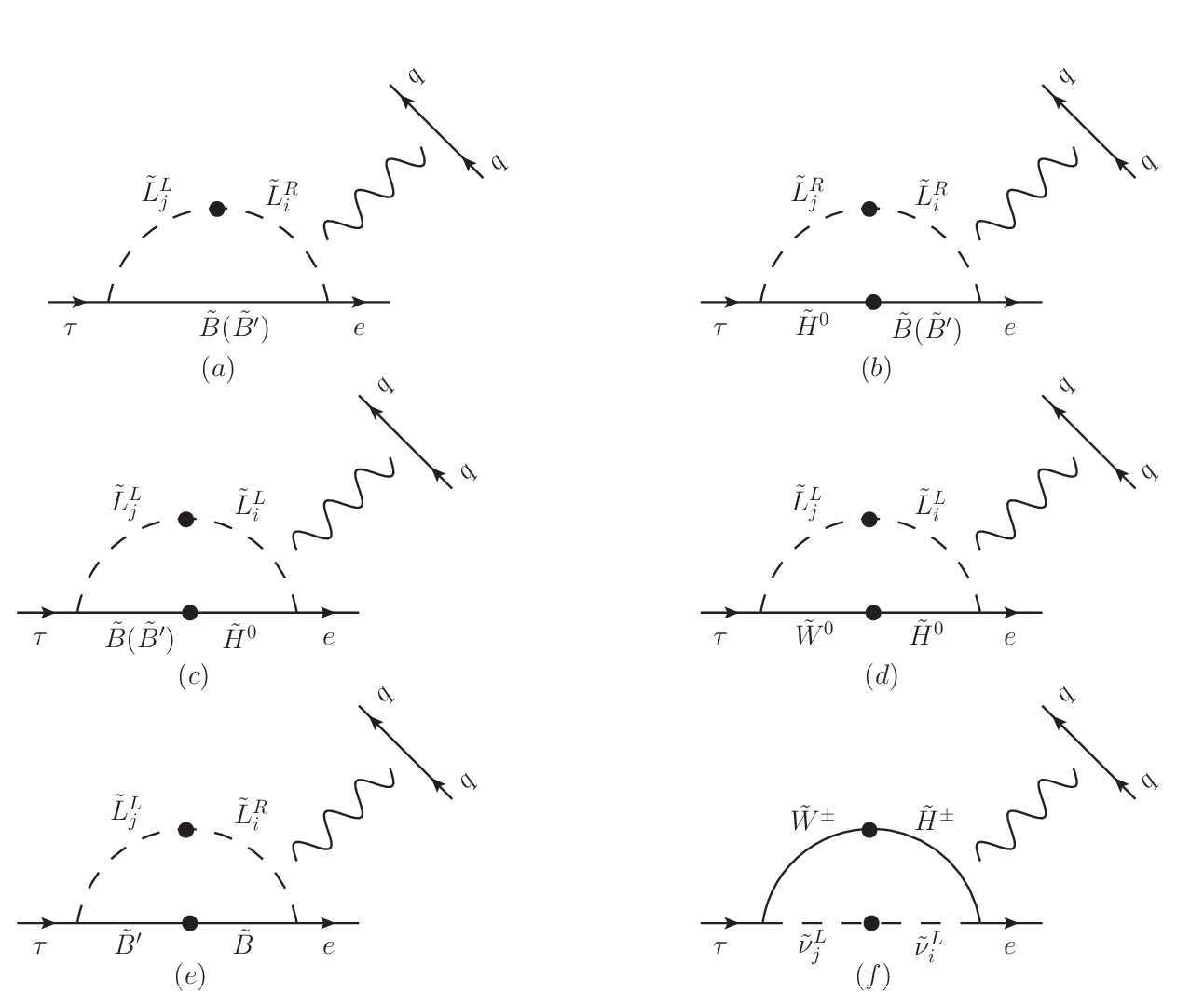}
\caption{Feynman diagrams for $\tau\rightarrow e M^+M^-$ in the MIA.}\label{MIA}
\end{figure}

Representative mass insertion diagrams for the process $\tau \to e M^+ M^-$ in the N-B-LSSM are presented in Fig.\ref {MIA}. The advantage of the mass insertion approximation (MIA) method lies in its ability to express flavor violation explicitly through mass insertions $\Delta_{ij}^{AB}(A,B=L,R)$ in the propagators. This enables us to write the LFV amplitudes directly in terms of the off-diagonal components of the soft SUSY breaking slepton mass matrices $m^2_{\tilde{L}}, m^2_{\tilde{E}}$ and the trilinear coupling matrix $T_e,~T_\nu$. With proper expansions, this approach leads to much simpler analytical expressions, allowing us to clearly identify the dominant contributions to LFV at the analytical level.

As an example, Fig.\ref {MIA}(a) shows a typical one-loop contribution mediated by $\tilde{B}$ and the slepton mass insertions between $\tilde{L}^L_j$ and $\tilde{L}^R_i$, with the amplitude given by:
\begin{eqnarray}
\mathcal{M}(\tilde{L}^L_j,\tilde{L}^R_i,\tilde{B})=\frac{-Q_q e^2}{k^2}\frac{M_1(m_e+m_\tau)}{\Lambda^2}\Delta^{LR}_{ij}g_1^2\mathcal{I}(x_1,x_{\tilde{L}^L_j},x_{\tilde{L}^R_i})
\times(\bar{e} \gamma^\mu P_L \tau)(\bar{q} \gamma_\mu P_L q+\bar{q} \gamma_\mu P_R q),
\end{eqnarray}
where the loop function $\mathcal{I}(x,y,z)$ is given by:
\begin{eqnarray}
&&\mathcal{I}(x,y,z)=\frac{1}{16\pi^2}\Big[\frac{x  \left(x^3-3 x y z+y z (y+z)\right)\ln x}{(x-y)^3
   (x-z)^3}+\frac{-3 x^2+x (y+z)+y z}{2 (x-y)^2 (x-z)^2}\nonumber\\&&~~~~~~~~~~~~-\frac{x y
   \ln y}{(x-y)^3 (y-z)}-\frac{x z \ln z}{(x-z)^3 (z-y)}\Big].
\end{eqnarray}

To better illustrate the parameter dependence, we consider a simplified scenario where all superpartner masses are nearly degenerate:
\begin{eqnarray}
&&M_1=m_{\tilde{L}_L}=m_{\tilde{L}_R}=M_{SUSY}.
\end{eqnarray}

In this degenerate limit, the loop function reduces to a constant $\mathcal{I}(1,1,1)=\frac{1}{192\pi^2}$, and the mass insertion can be expressed as:
\begin{eqnarray}
&&\Delta^{LR}_{ij}=m_{l_j}m_{\tilde{L}_L}\delta^{LR}_{ij}.
\end{eqnarray}

This clearly shows that the LFV amplitude is directly controlled by $\Delta^{LR}_{ij}$ including $T_{e_{ij}}$. In the similar way, the other MIA diagrams can also be analyzed, and we do not research them in detail anymore in this work. In the whole, the results depend on the off-diagonal elements of $m_{\tilde{L}, \tilde{E}}^2, T_e$ and $T_\nu$. Therefore, the simplified expressions clearly reveal the parametric dependence of the LFV amplitudes.

In summary, while the main body of this work is based on the mass eigenstate method to ensure precision and completeness, the inclusion of the MIA method provides an intuitive and analytical perspective on the dependence of LFV processes on key parameters. This supplementary analysis enhances the physical interpretability of the results and offers a useful framework for exploring other LFV processes in future studies.

\subsection{branching ratios}

Once the effective amplitudes at the quark level are determined, we can calculate the corresponding branching ratios \cite{Cirigliano}.
\begin{eqnarray}
&&\rm BR(\tau \to e \pi^+\pi^-)\simeq 1.9\times 10^{2}|\Gamma^e_{\gamma}|^2_{\tau e}+ 1.0\times10^{-8}|C_{GG}|^2+0.13|C^{ed}_{SRR}+C^{ed}_{SRL}|^2_{ss}\nonumber\\&&\hspace{3.2cm}
+ \Big(0.17\Big|[C^{eq}_{SRR} +C^{eq}_{SRL}]_{\tau e(qq)^{(0)}}\Big|^2
+ 0.5\Big|[C^{eq}_{VLL}+C^{eq}_{VLR}]_{\tau e(qq)^{(1)}}\Big|^2 \Big) \nonumber\\&&\hspace{3.2cm}
+1.0\Big|[ C^{ed}_{TRR}]_{\tau edd}-[C^{eu}_{ TRR} ]_{\tau euu}\Big|^2 .
\end{eqnarray}
with
\begin{eqnarray}
&&(\Gamma^e_{\gamma})_{\tau e}=\frac{v e m_\tau}{2 \sqrt{\alpha\pi}} A^R_2-\frac{16}{3}\Big(\frac{i\prod_{VT} (0)}{v}\Big)[C^{eu}_{TRR}]_{e \tau uu},~~~\nonumber\\&&
[C_{GG}]_{\tau e}=\frac{1}{3}\sum_{q=b,c}\frac{v}{m_q}[C^{eq}_{SRR} +C^{eq}_{SRL}]_{\tau eqq},~~(v=\sqrt{v_u^2+v_d^2},~\alpha\approx\frac{e^2}{\hbar c}).
\end{eqnarray}
where the non-perturbative parameter $(i\prod_{VT} (0)/v)\approx 1.6\times10^{-4}$ and the notation $(qq)^{(0),(1)}$ indicates that the  isoscalar or isovector ($uu \pm dd$) combination of Wilson coefficients has to be taken, $A^R_2$ can be found in Ref.\cite{SRZ}.
\begin{eqnarray}
&&\rm BR(\tau \to e \pi^+ K^-)\simeq 0.17 \Big|C^{ed}_{VLL}+C^{ed}_{VLR}\Big|^2_{\tau eds}+ 0.16\Big|C^{ed}_{SRR} +C^{ed}_{SRL}\Big|^2_{\tau eds}.
\end{eqnarray}
\begin{eqnarray}
&&\rm BR(\tau \rightarrow e K^+ K^-)=0.59\Big|\Big(C^{ed}_{VLL}+C^{ed}_{VLR} \Big)_{\tau ess}\Big|^2
+(1.0 \pm 0.1) \cdot 10^{-2} \Big|\Big(C^{eu}_{VLL}+C^{eu}_{VLR} \Big)_{\tau e uu}\Big|^2  \nonumber\\&&\hspace{3.5cm}
+ (0.6 \pm 0.2) \cdot 10^{-3} \Big|\Big(C^{ed}_{VLL}+C^{ed}_{VLR} \Big)_{\tau e dd}\Big|^2\nonumber\\&&\hspace{3.5cm}
-(4.6 \pm 0.2) \cdot 10^{-2}\Big(C^{ed}_{VLL}+C^{ed}_{ VLR} \Big)_{\tau e ss} \Big(C^{eu}_{VLL}+C^{eu}_{VLR} \Big)_{\tau e uu}
\nonumber\\&&\hspace{3.5cm}
-(4.3 \pm 1.5) \cdot 10^{-3}\Big(C^{ed}_{VLL}+C^{ed}_{VLR} \Big)_{\tau e ss} \Big(C^{ed}_{VLL}+C^{ed}_{VLR} \Big)_{\tau e dd}
\nonumber\\&&\hspace{3.5cm}
+ (3.5 \pm 0.8)\cdot 10^{-3} \Big(C^{ed}_{VLL}+C^{ed}_{VLR} \Big)_{\tau e dd} \Big(C^{eu}_{VLL}+C^{eu}_{VLR} \Big)_{\tau e uu}.
\end{eqnarray}

The final Wilson coefficients $C^{eq}_{VLL},~C^{eq}_{VLR},~C^{eq}_{SRR},~C^{eq}_{SRL}$ and $C^{eq}_{TRR}$ are obtained from the sum of these diagrams' amplitudes. The corresponding effective operators are $(\bar e{\gamma^\mu }{P_L}\tau)(\bar q{\gamma_\mu}{P_L}q),~(\bar e{\gamma^\mu}{P_L}\tau)(\bar q{\gamma_\mu }{P_R} q),~(\bar e P_R\tau)(\bar q P_R q),~(\bar e P_R \tau)(\bar q P_L q),~(\bar e\sigma^{\mu\nu}{P_R}\tau)(\bar q{\sigma_{\mu\nu}}{P_R}q)$, respectively. For convenience, the final Wilson coefficients are analyzed in the generic form, which can simplify the work. Taking $C^{eq}_{VLL}$ as an example, and see the appendix \ref{A1} for the rest.
\begin{eqnarray}
&&C^{eq}_{VLL}=\sum_{F_1,F_2=\chi^\pm,\chi^\pm} \sum_{S_1,S_2=\tilde{\nu},\tilde{u}}\frac{1}{4}I_8(x_{F_1},x_{F_2},x_{S_1},x_{S_2})H_L^{S_1\tau\bar{F_1}}H_R^{S_1 F_2\bar{e}}H_R^{S_2 F_1 \bar{d}}H_L^{S_2 d\bar{F_2}}\nonumber\\&&~~~~~~~~
-\sum_{F_1,F_2=\chi^0,\chi^0} \sum_{S_1,S_2=\tilde{e},\tilde{q}}\frac{1}{2}\Big[I_9(x_{F_1},x_{F_2},x_{S_1},x_{S_2})H_L^{S^*_1\tau\bar{F_1}}H_R^{S_1 F_2\bar{e}}H_R^{S_2 F_2 \bar{q}}H_L^{S^*_2 q \bar{F_1 }}\nonumber\\&&~~~~~~~~
-\frac{1}{2}I_8(x_{F_1},x_{F_2},x_{S_1},x_{S_2})H_L^{S^*_1\tau\bar{F_1}}H_R^{S_1 F_2\bar{e}}H_R^{S_2 F_1 \bar{q}}H_L^{S^*_2 q\bar{F_2}}\Big]\nonumber\\&&~~~~~~~~
+\sum_{F=\chi^0,\chi^{\pm}} \sum_{S=\tilde{e},\tilde{\nu}}\frac{1}{m_Z^{2}}\Big[\frac{1}{2}I_5(x_F,x_{S_1},x_{S_2})[H_R^{S_2 F \bar{e}}H_L^{S^* \tau \bar{F}}H^{Z S_1 S^*_1}H_L^{\bar{q} Z q}\nonumber\\&&~~~~~~~~
-H_R^{S F_2 \bar{e}}H_R^{Z F_1 \bar{F_2}}H_L^{S^* \tau \bar{F_1}}H_L^{\bar{q} Z q}]
+m_{F_1}m_{F_2}I_6(x_S,x_{F_2},x_{F_1})H_R^{S F_2 \bar{e}}H_L^{Z F_1 \bar{F_2}}H_L^{S^* \tau \bar{F_1}}H_L^{\bar{q} Z q}\Big]\nonumber\\&&~~~~~~~~
+\frac{1}{(m^2_e-m^2_\tau)}\Big[\frac{1}{2}I_1(x_F,x_S)(m^2_e-m^2_\tau ) [\frac{Q_q e^2}{k^2}H_R^{S^* F \bar{e}} H_L^{S \tau \bar{F}}\nonumber\\&&~~~~~~~~-\frac{1}{(k^2-m^2_Z)}H_R^{S^* F \bar{e}} H_L^{S \tau \bar{F}}H_L^{Z \tau \bar{e}}H_L^{\bar{q} Z q}]
+(m^2_e-m^2_\tau)[I_3(x_F,x_S)-I_4(x_F,x_S)]\nonumber\\&&~~~~~~~~\times[\frac{Q_q e^2}{k^2}m_F(m_e  H_L^{S^* F \bar{e}} H_L^{S \tau \bar{F}}+m_\tau H_R^{S^* F \bar{e}} H_R^{S \tau \bar{F}})-\frac{1}{(k^2-m^2_Z)}\nonumber\\&&~~~~~~~~\times m_F(m_e  H_L^{S^* F \bar{e}} H_L^{S \tau \bar{F}}H_L^{Z \tau \bar{e}}H_L^{\bar{q} Z q}+m_\tau H_R^{S^* F \bar{e}} H_R^{S \tau \bar{F}}H_L^{Z \tau \bar{e}}H_L^{\bar{q} Z q})]\Big]\nonumber\\&&~~~~~~~~
+\sum_{F=\chi^0} \sum_{S=\tilde{e}}\frac{-Q_q e^2}{k^2}\Big[\frac{1}{2}I_1(x_F,x_S) H_R^{S F \bar{e}} H_L^
{S^* \tau \bar{F}} +[I_2(x_F,x_S)-I_4(x_F,x_S)]\nonumber\\&&~~~~~~~~
\times[(m^2_\tau+m^2_e)H_R^{S F \bar{e}} H_L^{S^* \tau \bar{F}} +m_e m_\tau H_L^{S F \bar{e}} H_R^{S^* \tau \bar{F}} ]\nonumber\\&&~~~~~~~~+[I_2(x_F,x_S)-I_3(x_F,x_S)]
[m_F (m_e H_L^{S F \bar{e}} H_L^{S^* \tau \bar{F}} + m_\tau H_R^{S F \bar{e}} H_R^{S^* \tau \bar{F}} )]\Big]\nonumber\\&&~~~~~~~~
+\sum_{F=\chi^{\pm}} \sum_{S=\tilde{\nu}}\frac{Q_q e^2}{k^2}\Big[[\frac{1}{2}I_1(x_F,x_S)-m^2_F I_3(x_F,x_S)] H_R^{S F \bar{e}} H_L^{S^* \tau \bar{F}}\nonumber\\&&~~~~~~~~+[2I_4(x_F,x_S)-I_2(x_F,x_S)-I_3(x_F,x_S)]
[(m^2_\tau+m^2_e)H_R^{S F \bar{e}}  H_L^{S^* \tau \bar{F}} \nonumber\\&&~~~~~~~~+m_e m_\tau H_L^{S F \bar{e}}  H_R^{S^* \tau \bar{F}} ]+[I_4(x_F,x_S)-I_3(x_F,x_S)]
[m_F (m_e H_L^{S F \bar{e}}  H_L^{S^* \tau \bar{F}} \nonumber\\&&~~~~~~~~+m_\tau H_R^{S F \bar{e}}  H_R^{S^* \tau \bar{F}} )]\Big].
\end{eqnarray}

\section{numerical results}

In this section, we perform a numerical analysis of LFV processes and systematically investigate the model parameters under current experimental constraints. To obtain reasonable numerical results, several sensitive parameters are explored, and the processes $\tau \rightarrow e \pi^+\pi^-$, $\tau\rightarrow e\pi^+K^-$ and $\tau \rightarrow e K^+ K^-$ are discussed in detail in three subsections. Notably, since the experimental upper limit on the $\tau \rightarrow e\gamma$ process imposes the most stringent constraints on the parameter space of the N-B-LSSM, its impact on LFV must be thoroughly taken into account \cite{SRZ}. Furthermore, we adopt the lightest CP-even Higgs mass $m_{h^0}=125.20 \pm 0.11~{\rm GeV}$ \cite{pdg}. For the mass of the added heavy vector boson $Z^\prime$, the latest experimental constraint is $M_{Z^\prime}>5.1~{\rm TeV}$, a significantly stronger bound than previous limits \cite{ATLAS:2019erb}. Given that $M_{Z^\prime}$ is much larger than $M_Z$, its contribution to the amplitude is negligible and thus not calculated in this work. The lower limit on the ratio $M_{Z^\prime}$/$g_B$ is set to $6~{\rm TeV}$ at $99\%$ C.L. \cite{Cacciapaglia:2006pk,Carena:2004xs}. Considering constraints from LHC data \cite{wx1,wx2,wx3,wx4,wx5,wx6,wx7,TanBP}, we set the following parameter conditions: the slepton mass greater than $700~{\rm GeV}$, the chargino mass greater than $1100~{\rm GeV}$, and the squark mass greater than $1600~{\rm GeV}$, with the experimental value of $\tan \beta_\eta$ being less than 1.5. In addition, the constraints of Charge and Color Breaking (CCB) are also taken into account \cite{HAN1,HAN2}. Based on these rigorous experimental requirements, we collect extensive data, and the relationships among various parameters are graphically illustrated. Through a systematic analysis of these plots and the experimental upper limits on the branching ratios, we identify a viable parameter space that explains LFV phenomena.

Considering the above constraints in the front paragraph, we use the following parameters in the N-B-LSSM:
\begin{eqnarray}
&&\tan{\beta}_{\eta}=0.9,~~Y_{Xii}=0.5,~~T_{\lambda}=1~{\rm TeV},~~T_{\lambda_2}=1~{\rm TeV},~~T_{\kappa}=-2.5~{\rm TeV},\nonumber\\
&&T_{uii}=1~{\rm TeV},~~T_{dii}=1~{\rm TeV},~~T_{Xii}=-4~{\rm TeV},~~M_1=0.4~{\rm TeV},~~M_2=1.2~{\rm TeV},\nonumber\\
&&M_{\tilde{\nu} ii}^2=2.5~{\rm TeV^2},M_{\tilde{Q}ii}^2=3.1~{\rm TeV^2},M_{\tilde{U}ii}^2=2.2~{\rm TeV^2},M_{\tilde{D}ii}^2=2.8~{\rm TeV^2},(i=1,2,3).
\end{eqnarray}

To simplify the numerical discussion, we employ the parameter relationships and analyze their variations in numerical analysis:
\begin{eqnarray}
&&\tan\beta,~~g_B,~~g_{YB},~~\lambda,~~\lambda_2,~~v_S,~~\kappa,~~M_{BL},~~M_{BB'},\nonumber\\
&&M_{\tilde{L}ii}^2=M_{\tilde{L}}^2,~~M_{\tilde{E}ii}^2=M_{\tilde{E}}^2,~~M_{\tilde{L}ij}^2=M_{\tilde{L}ji}^2,~~M_{\tilde{E}ij}^2=M_{\tilde{E}ji}^2,\nonumber\\
&&T_{eii}=T_{e},~~T_{\nu ii}=T_{\nu},~~T_{eij}=T_{eji},~~T_{\nu ij}=T_{\nu ji},(i,j=1,2,3,~i\neq j).
\end{eqnarray}

If we do not especially declare, the off-diagonal elements of the used parameters are assumed to be zero.

In the framework of the N-B-LSSM, LFV originates primarily from the flavor off-diagonal structures introduced by soft SUSY breaking terms, which violate lepton flavor conservation. At the loop diagram level, these off-diagonal parameters alter the mass eigenstates and interaction vertices of internal SUSY particles, thereby inducing $\tau \to e$ transitions at the effective vertex. The relevant parameters include: (i) the off-diagonal element $M^2_{\tilde{L}13}$ in the left-handed slepton mass matrix, also appearing in the CP-even and CP-odd sneutrino sectors, which induces flavor mixing between $\tilde{e}_L$ and $\tilde{\tau}_L$, significantly contributing to both $\tilde{e}-\chi^0$ and $\tilde{\nu}^{R,I}-\chi^\pm$ loop diagrams; (ii) the off-diagonal element $M^2_{\tilde{E}13}$ in the right-handed slepton mass matrix, responsible for the mixing between $\tilde{e}_R$ and $\tilde{\tau}_R$, which mainly affects the flavor structure of $\tilde{e}-\chi^0$ loop diagrams; (iii) the off-diagonal term $M^2_{\tilde{\nu}13}$ in the sneutrino mass matrix, which influences the masses and mixings of CP-even and CP-odd sneutrinos. It plays a crucial role in $\tilde{\nu}^{R,I}-\chi^\pm$ loop diagrams; (iv) the trilinear coupling $T_{e13}$, which enhances the couplings between different slepton flavors such as $\tilde{e}_L$ and $\tilde{\tau}_R$, contributing to $\tilde{e}-\chi^0$ loop diagrams; and (v) the sneutrino trilinear coupling $T_{\nu13}$, which affects the mass spectra and mixings of CP-even and CP-odd sneutrinos; It significantly contributes to $\tilde{\nu}^{R,I}-\chi^\pm$ loop diagrams and is a key parameter determining the LFV transition rates. The effective couplings in the loop diagrams are directly governed by the aforementioned flavor-violating parameters. As such, these parameters collectively determine both the LFV transition mechanisms and the resulting branching ratios. In fact, in addition to the soft-breaking parameters, given the non-zero masses of neutrinos, the CKM-like matrix in the lepton sector is another source of the flavor violation. This source, however, is not important for the processes due to the small mass splitting of neutrinos (like GIM mechanism).

\subsection{The process of $\tau \rightarrow e \pi^+\pi^-$}

\begin{figure}[ht]
\setlength{\unitlength}{5mm}
\centering
\includegraphics[width=2.9in]{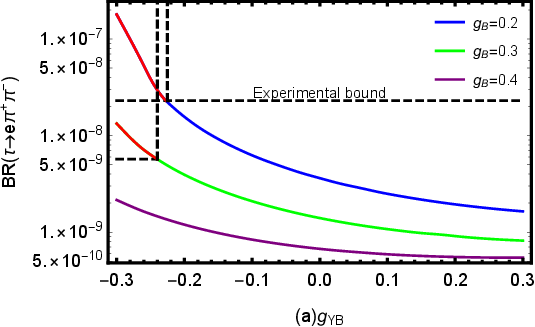}
\setlength{\unitlength}{5mm}
\centering
\includegraphics[width=2.9in]{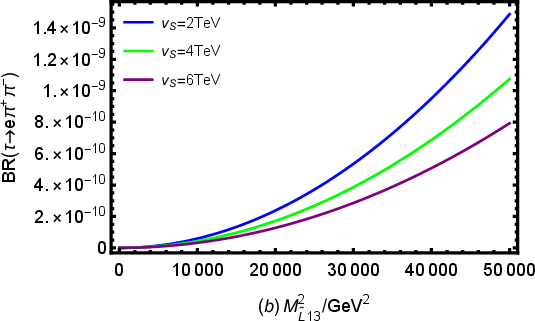}
\setlength{\unitlength}{5mm}
\centering
\includegraphics[width=2.9in]{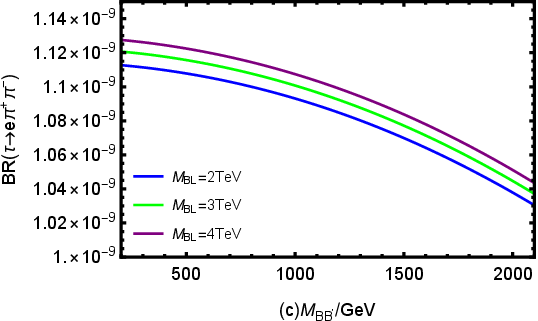}
\setlength{\unitlength}{5mm}
\centering
\includegraphics[width=2.9in]{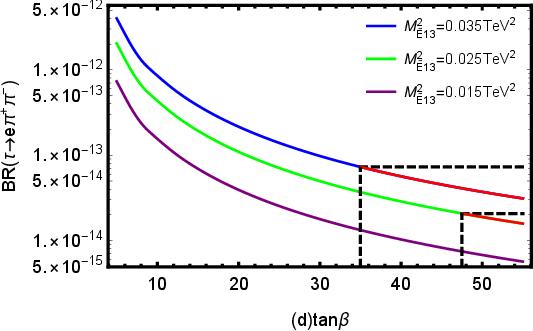}
\setlength{\unitlength}{5mm}
\centering
\includegraphics[width=2.9in]{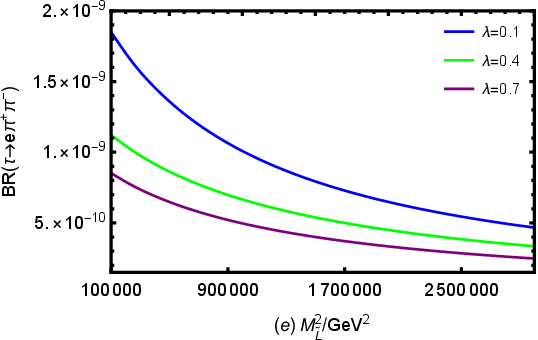}
\caption{The influence of various parameters on BR($\tau \rightarrow e \pi^+\pi^-$).}\label{tauepipi}
\end{figure}

In the case of parameters $\lambda_2=-0.25,~\kappa=0.1,~T_{\nu}=1~{\rm TeV},~T_{e}=1.5~{\rm TeV},~M_{\tilde{E}}^2=1.7~{\rm TeV^2}$, we draw BR($\tau \rightarrow e \pi^+\pi^-$) diagrams under the influence of different parameters in Fig.\ref{tauepipi}.

Using the parameters $\lambda=0.4$,~$\tan\beta=25$,~$M_{BB'}=0.1~{\rm TeV}$,~$M_{BL}=1~{\rm TeV}$,~$v_S=4~{\rm TeV}$,~$M_{\tilde{L}13}^2=0.05~{\rm TeV^2}$,~$M_{\tilde{L}}^2=0.16~{\rm TeV^2}$, we plot BR($\tau \rightarrow e \pi^+\pi^-$) versus $g_{YB}$ in Fig.\ref{tauepipi}(a), where the blue, green and purple curves correspond to $g_B=0.2,~0.3,~0.4$ respectively. It can be clearly seen that for any given $g_B$, the branching ratio decreases monotonically with the increase of $g_{YB}$; meanwhile, under the same $g_{YB}$ condition, the larger the $g_B$, the smaller the branching ratio. Specifically, the top curve has successively exceeded the experimental upper limits of $\tau\rightarrow e\gamma$ and $\tau \rightarrow e \pi^+\pi^-$ in the region of $-0.3<g_{YB}<-0.22$. The middle curve only exceeds the limit of $\tau\rightarrow e\gamma$ in the region of $-0.3<g_{YB}<-0.24$, while the bottom curve remains below the existing experimental constraints throughout the entire scanned range. $g_B$ is the $U(1)_{B-L}$ gauge coupling constant. The mass matrices of several particles (neutralino,~slepton,~CP-even sneutrino,~CP-odd sneutrino,~up-squark,~down-squark,~CP-even Higgs) all have the important parameter $g_B$, which can improve the NP effect. $g_{YB}$ is the coupling constant for gauge mixing of $U(1)_Y$ and $U(1)_{B-L}$, which is a new parameter beyond MSSM and can bring new effect. The formation of this trend is mainly attributed to the dual role of the two coupling constants on the NP effect. On the one hand, both $g_B$ and $g_{YB}$ participate in the mass matrices and vertex structures of various SUSY particles, which can enhance LFV vertex coupling strengths. On the other hand, these two parameters simultaneously increase the mass of the particles and enhance the mass suppression effect of the loop propagator. The inhibition effect brought about by the improvement in mass significantly exceeds the amplitude increase caused by the coupling enhancement, resulting in the overall branching ratio decreasing as $g_B$ and $g_{YB}$ increased. Therefore, $g_B$ and $g_{YB}$ can be regarded as sensitive and critical parameters.

In the case of $g_{YB}=0.1$,~$g_B=0.3$,~$\lambda=0.4$,~$\tan\beta=25$,~$M_{BB'}=0.1~{\rm TeV}$,~$M_{BL}=1~{\rm TeV}$,~$M_{\tilde{L}}^2=0.16~{\rm TeV^2}$, Fig.\ref{tauepipi}(b) shows the trend of BR($\tau \rightarrow e \pi^+\pi^-$) changing with $M_{\tilde{L}13}^2$. Each curve corresponds to different $v_S$ values (the blue curve is $v_S=2~{\rm TeV}$, the green curve is $v_S=4~{\rm TeV}$, and the purple curve is $v_S=6~{\rm TeV}$). $M_{\tilde{L}13}^2$ is a off-diagonal element in the left-handed slepton mass matrix. Its non-zero value introduces lepton flavor mixing between the first generation ($e$) and the third generation ($\tau$). This flavor mixing makes flavor conservation less stringent, allowing LFV processes like $\tau \rightarrow e \pi^+\pi^-$ to occur. Therefore, the LFV effect is enhanced as $M_{\tilde{L}13}^2$ increases, causing the branching ratio to rise rapidly, and the three curves show obvious nonlinear positive growth. Comparing different $v_S$ values, a larger $v_S$ corresponds to a smaller branching ratio under the same $M_{\tilde{L}13}^2$ condition. $v_S$ is VEV of the Higgs singlet S and appears in almost all mass matrices involving spontaneous breaking of $U(1)_{B-L}$. In one-loop diagrams, the related particles participate in the propagation as internal propagators, which cause a significant propagation suppression effect when their masses increase with $v_S$, leading to a reduction in the branching ratio. In the numerical scan, it can be seen that the sensitivity of BR($\tau \rightarrow e \pi^+\pi^-$) to $M_{\tilde{L}13}^2$ is much higher than that of $v_S$.

Assuming $g_{YB}=0.1$,~$g_B=0.3$,~$\lambda=0.4$,~$\tan\beta=25$,~$v_S=4~{\rm TeV}$,~$M_{\tilde{L}13}^2=0.05~{\rm TeV^2}$,~$M_{\tilde{L}}^2=0.16~{\rm TeV^2}$, we show BR($\tau \rightarrow e \pi^+\pi^-$) varying with $M_{BB'}$ by the blue curve ($M_{BL}=2~{\rm TeV}$), green curve ($M_{BL}=3~{\rm TeV}$) and purple curve ($M_{BL}=4~{\rm TeV}$) in Fig.\ref{tauepipi}(c). The parameter $M_{BB'}$ represents the mass of the $U(1)_Y$ and $U(1)_{B-L}$ gaugino mixing, $M_{BL}$ is the mass of the superpartner for the gauge boson under the $U(1)_{B-L}$ group. $M_{BB'}$ and $M_{BL}$ present in the mass matrix of neutralino. Both of them affect the mixing and mass structure of neutralino, thereby influencing the amplitude of the process in the Feynman diagrams involving neutralino. As $M_{BB'}$ increases, the branching ratios of all curves decrease gradually. For any fixed $M_{BB'}$, a larger $M_{BL}$ results in  a slightly higher branching ratio. However, it can be seen from the variation of the vertical axis in Fig.\ref{tauepipi}(c) that the curves corresponding to the three different $M_{BL}$ values are very close to each other, and the range of changes in the vertical axis is relatively small. This indicates that the influence of both parameters on the branching ratio does exist, but the overall contribution is weak and belongs to the secondary correction. Therefore, in the parameter sensitivity analysis of the process, the influence of $M_{BB'}$ and $M_{BL}$ can be regard as a mild regulatory effect rather than a decisive factor.

Under the conditions $g_{YB}=0.1$,~$g_B=0.3$,~$\lambda=0.4$,~$M_{BB'}=0.1~{\rm TeV}$,~$M_{BL}=1~{\rm TeV}$,~$v_S=4~{\rm TeV}$,~$M_{\tilde{L}}^2=0.16~{\rm TeV^2}$, we study the relationship between BR($\tau \rightarrow e \pi^+\pi^-$) and $\tan\beta$ in Fig.\ref{tauepipi}(d), the curves are divided into three cases corresponding to $M_{\tilde{E}13}^2=0.035~{\rm TeV^2}$ (blue curve), $M_{\tilde{E}13}^2=0.025~{\rm TeV^2}$ (green curve) and $M_{\tilde{E}13}^2=0.015~{\rm TeV^2}$ (purple curve). The parameter $\tan\beta$ is defined as the ratio of the VEVs of the two Higgs doublets, namely $\tan\beta=v_u/v_d$. It influences the vertex couplings and particle masses by directly affecting $v_d$ and $v_u$. Since $\tan\beta$ appears in almost all the mass matrices of Fermions, scalars and Majoranas, it must be a highly sensitive parameter. For each curve, the branching ratio decreases by about 2 to 3 orders of magnitude as $\tan\beta$ increases from 5 to 55. $M_{\tilde{E}13}^2$ denotes the flavor off-diagonal term between $\tau$ and $e$ in the slepton softbreaking mass matrix, which essentially reflects the mixing strength among right-handed slepton SUSY particles. A larger $M_{\tilde{E}13}^2$ implies stronger lepton flavor mixing, which amplifies the loop contributions, leading to an upward shift in the overall branching ratio level. The right ends of the blue and green curves are marked in red, indicating that the corresponding parameter points have not exceeded the current experimental upper limit of $\tau \rightarrow e \pi^+\pi^-$, they have violated the constraint of $\tau\rightarrow e\gamma$. The purple curve is allowed under both experimental limits due to its relatively low overall branching ratio.

Based on $g_{YB}=0.1$,~$g_B=0.3$,~$\tan\beta=25$,~$M_{BB'}=0.1~{\rm TeV}$,~$M_{BL}=1~{\rm TeV}$,~$v_S=4~{\rm TeV}$,~$M_{\tilde{L}13}^2=0.05~{\rm TeV^2}$, Fig.\ref{tauepipi}(e) illustrates the trend of BR($\tau \rightarrow e \pi^+\pi^-$) as $M_{\tilde{L}}^2$ varies, with three curves plotted corresponding to $\lambda=0.1$ (blue curve), $\lambda=0.4$ (green line) and $\lambda=0.7$ (purple line). $M_{\tilde{L}}^2$ represents the diagonal term in the mass matrices of slepton as well as CP-even sneutrino and CP-old sneutrino, whose values affect the overall mass scale of the new physical particles involved in the loop process. With the increases of $M_{\tilde{L}}^2$, the branching ratio shows a gradual downward trend. The higher $M_{\tilde{L}}^2$ means that the greater the mass of the relevant virtual particles, resulting in the significant suppression of the contribution of these particles in the low-energy process. In the superpotential, the term $\lambda\hat{S}\hat{H}_u\hat{H}_d$ involves the coupling constant $\lambda$. For a fixed $M_{\tilde{L}}^2$, the smaller $\lambda$ value corresponds to a larger branching ratio. Although both have inhibitory effects on the branching ratio, judging from the inclination of the curves in Fig.\ref{tauepipi}(e), BR($\tau \rightarrow e \pi^+\pi^-$) is more sensitive to the change of $M_{\tilde{L}}^2$. This is because the variation of the branching ratio with $M_{\tilde{L}}^2$ under the same $\lambda$ is much greater than the difference under different $\lambda$ at a fixed $M_{\tilde{L}}^2$. This suggests that $M_{\tilde{L}}^2$ is a more critical parameter than $\lambda$ in parameter constraints and sensitivity analysis.

\subsection{The process of $\tau \rightarrow e \pi^+K^-$}

In order to better explain how variables affect the branching ratio of ($\tau \rightarrow e \pi^+K^-$), we randomly scan the parameters. All the parameters involved are expressed in tabular form.

\begin{table*}
\caption{Scanning parameters for Fig.{\ref {tauepik1}}}\label{a}
\begin{tabular}{|c|c|c|}
\hline
Parameters&Min&Max\\
\hline
$\hspace{1.5cm}\lambda\hspace{1.5cm}$ &$\hspace{1.5cm}0.05\hspace{1.5cm}$& $\hspace{1.5cm}0.4\hspace{1.5cm}$\\
\hline
$\hspace{1.5cm}\lambda_2\hspace{1.5cm}$ &$\hspace{1.5cm}-0.3\hspace{1.5cm}$& $\hspace{1.5cm}-0.05\hspace{1.5cm}$\\
\hline
$\hspace{1.5cm}\kappa\hspace{1.5cm}$ &$\hspace{1.5cm}0.01\hspace{1.5cm}$& $\hspace{1.5cm}0.7\hspace{1.5cm}$\\
\hline
$\hspace{1.5cm}g_{YB}\hspace{1.5cm}$ &$\hspace{1.5cm}-0.4\hspace{1.5cm}$ &$\hspace{1.5cm}0.2\hspace{1.5cm}$\\
\hline
$\hspace{1.5cm}g_B\hspace{1.5cm}$ & $\hspace{1.5cm}0.3\hspace{1.5cm}$ &$\hspace{1.5cm}0.8\hspace{1.5cm}$\\
\hline
$\hspace{1.5cm}\tan\beta\hspace{1.5cm}$ & $\hspace{1.5cm}5\hspace{1.5cm}$ &$\hspace{1.5cm}50\hspace{1.5cm}$\\
\hline
$\hspace{1.5cm}v_S/\rm GeV\hspace{1.5cm}$ & $\hspace{1.5cm}2000\hspace{1.5cm}$ &$\hspace{1.5cm}7000\hspace{1.5cm}$\\
\hline
$\hspace{1.5cm}M_{BB'}/\rm GeV\hspace{1.5cm}$ & $\hspace{1.5cm}100\hspace{1.5cm}$ &$\hspace{1.5cm}3000\hspace{1.5cm}$\\
\hline
$\hspace{1.5cm}M_{BL}/\rm GeV\hspace{1.5cm}$ & $\hspace{1.5cm}500\hspace{1.5cm}$ &$\hspace{1.5cm}5000\hspace{1.5cm}$\\
\hline
$\hspace{1.5cm}M^2_{\tilde{L}13}/\rm GeV^2\hspace{1.5cm}$ &$\hspace{1.5cm}0\hspace{1.5cm}$& $\hspace{1.5cm}10^5\hspace{1.5cm}$\\
\hline
$\hspace{1.5cm}M^2_{\tilde{E}}/\rm GeV^2\hspace{1.5cm}$ &$\hspace{1.5cm}2\times10^{5}\hspace{1.5cm}$& $\hspace{1.5cm}3\times10^6\hspace{1.5cm}$\\
\hline
\end{tabular}
\end{table*}

\begin{figure}[ht]
\setlength{\unitlength}{5mm}
\centering
\includegraphics[width=2.9in]{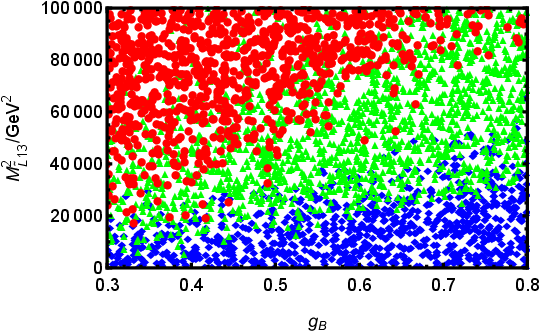}
\caption{Under the premises of current limits on LFV decays $\tau \rightarrow e \pi^+K^-$ and $\tau \rightarrow e \gamma$, reasonable parameter space is selected to scatter points. $\textcolor{blue}{\blacklozenge}~(0<\rm BR(\tau \rightarrow e \pi^+K^-)<3\times 10^{-11}),~\textcolor{green}{\blacktriangle}~(3\times 10^{-11}\leq \rm BR(\tau \rightarrow e \pi^+K^-)<3\times 10^{-10}),~ \textcolor{red}{\bullet}~(3\times 10^{-10}\leq \rm BR(\tau \rightarrow e \pi^+K^-)<3.7\times 10^{-8})$.}\label{tauepik1}
\end{figure}

Supposing the parameters with $T_\nu=1~{\rm TeV},~T_e=1.5~{\rm TeV}$ and $M_{\tilde{L}}^2=0.16~{\rm TeV^2}$, the relationship between $g_B$ and $M_{\tilde{L}13}^2$ is shown in Fig.\ref {tauepik1}. Fig.\ref{tauepik1} is obtained from the parameters shown in the Table \ref{a}. $\textcolor{blue}{\blacklozenge}$ are mainly distributed in the lower area of Fig.\ref {tauepik1}, especially concentrated in the lower right corner. Within the range of $M_{\tilde{L}13}^2<2\times10^4~{\rm GeV^2}$, $\textcolor{blue}{\blacklozenge}$ are most densely distributed regardless of the value of $g_B$. $\textcolor{red}{\bullet}$ are concentrated in the upper left part of Fig.\ref {tauepik1}, which are particularly dense in the region of $0.3<g_B<0.6,~4\times10^4~{\rm GeV^2}<M_{\tilde{L}13}^2<1\times10^5~{\rm GeV^2}$, and there are basically no $\textcolor{blue}{\blacklozenge}$ and $\textcolor{green}{\blacktriangle}$ inside the area, indicating that most of the points in this parameter range correspond to larger branching ratios. $\textcolor{green}{\blacktriangle}$ are located between $\textcolor{red}{\bullet}$ and $\textcolor{blue}{\blacklozenge}$ areas. $\textcolor{green}{\blacktriangle}$ gradually transition to the $\textcolor{red}{\bullet}$ area towards the upper left and connect to the $\textcolor{blue}{\blacklozenge}$ area towards the lower right, presenting a boundary structure along the diagonal direction. This distribution indicates that the two parameters have a significant joint effect on BR($\tau \rightarrow e \pi^+K^-$), the branching ratio tends to be larger for larger $M_{\tilde{L}13}^2$ and smaller $g_B$.

\begin{figure}[ht]
\setlength{\unitlength}{5mm}
\centering
\includegraphics[width=2.9in]{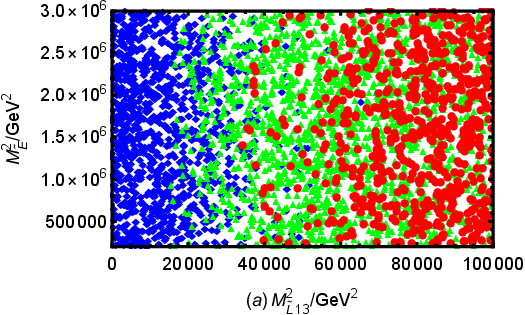}
\setlength{\unitlength}{5mm}
\centering
\includegraphics[width=2.7in]{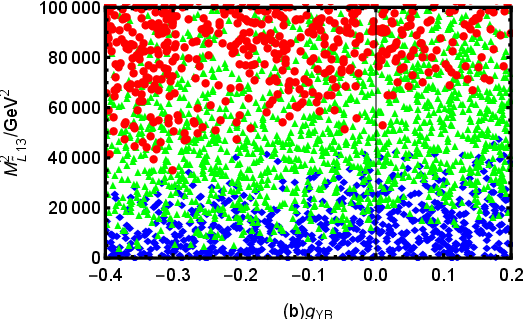}
\caption{Under the premises of current limits on LFV decays $\tau \rightarrow e \pi^+K^-$ and $\tau \rightarrow e \gamma$, reasonable parameter space is selected to scatter points. $\textcolor{blue}{\blacklozenge}~(0<\rm BR(\tau \rightarrow e \pi^+K^-)<3\times 10^{-11}),~\textcolor{green}{\blacktriangle}~(3\times 10^{-11}\leq \rm BR(\tau \rightarrow e \pi^+K^-)<3\times 10^{-10}),~ \textcolor{red}{\bullet}~(3\times 10^{-10}\leq \rm BR(\tau \rightarrow e \pi^+K^-)<3.7\times 10^{-8})$.}\label{tauepik24}
\end{figure}

Next, we scatter points on $\tau \rightarrow e \pi^+K^-$ in Fig.\ref{tauepik24} with the parameters in the Table \ref{bd}. Fig.\ref {tauepik24}(a) shows the distribution of BR($\tau \rightarrow e \pi^+K^-$) as the parameters $M_{\tilde{L}13}^2$ and $M_{\tilde{E}}^2$ change. It can be observed that $\textcolor{blue}{\blacklozenge}$~$\textcolor{green}{\blacktriangle}$ and $\textcolor{red}{\bullet}$ uniformly cover the entire scanning interval in the $M_{\tilde{E}}^2$ direction, and the trend of change is not significant, indicating that the impact of $M_{\tilde{E}}^2$ on the branching ratio is very small. In the direction of the horizontal axis, with the variation of $M_{\tilde{L}13}^2$, Fig.\ref {tauepik24}(a) shows an extremely obvious colour partition. In the $M_{\tilde{L}13}^2<20000~{\rm GeV^2}$ interval, almost all the data points are $\textcolor{blue}{\blacklozenge}$, indicating that the branching ratio is in the lowest order of magnitude when this mixing parameter is relatively small. As $M_{\tilde{L}13}^2$ increases to the $2\times 10^4~{\rm GeV^2} \sim 6\times 10^4~{\rm GeV^2}$ region, $\textcolor{green}{\blacktriangle}$ start to appear in large numbers, while $\textcolor{blue}{\blacklozenge}$ decrease significantly, and the branching ratio rises rapidly to the middle order of magnitude. When $M_{\tilde{L}13}^2>6\times 10^4~{\rm GeV^2}$, $\textcolor{red}{\bullet}$ become the dominant and $\textcolor{blue}{\blacklozenge}$~$\textcolor{green}{\blacktriangle}$ basically disappear, indicating that the branching ratio in this area is significantly enhanced and close to the experimental limit. The change of $M_{\tilde{L}13}^2$ is a sensitive factor determining the order-of-magnitude transition in branching ratio, which is consistent with what is reflected in Fig.\ref {tauepik1}.

\begin{table*}
\caption{Scanning parameters for Fig.{\ref {tauepik24}} and Fig.{\ref {tauekk1}}}\label{bd}
\begin{tabular}{|c|c|c|}
\hline
Parameters&Min&Max\\
\hline
$\hspace{1.5cm}\lambda\hspace{1.5cm}$ &$\hspace{1.5cm}0.05\hspace{1.5cm}$& $\hspace{1.5cm}0.4\hspace{1.5cm}$\\
\hline
$\hspace{1.5cm}\lambda_2\hspace{1.5cm}$ &$\hspace{1.5cm}-0.3\hspace{1.5cm}$& $\hspace{1.5cm}-0.05\hspace{1.5cm}$\\
\hline
$\hspace{1.5cm}\kappa\hspace{1.5cm}$ &$\hspace{1.5cm}0.01\hspace{1.5cm}$& $\hspace{1.5cm}0.7\hspace{1.5cm}$\\
\hline
$\hspace{1.5cm}g_{YB}\hspace{1.5cm}$ &$\hspace{1.5cm}-0.4\hspace{1.5cm}$ &$\hspace{1.5cm}0.2\hspace{1.5cm}$\\
\hline
$\hspace{1.5cm}g_B\hspace{1.5cm}$ & $\hspace{1.5cm}0.3\hspace{1.5cm}$ &$\hspace{1.5cm}0.8\hspace{1.5cm}$\\
\hline
$\hspace{1.5cm}\tan\beta\hspace{1.5cm}$ & $\hspace{1.5cm}5\hspace{1.5cm}$ &$\hspace{1.5cm}50\hspace{1.5cm}$\\
\hline
$\hspace{1.5cm}v_S/\rm GeV\hspace{1.5cm}$ & $\hspace{1.5cm}2000\hspace{1.5cm}$ &$\hspace{1.5cm}7000\hspace{1.5cm}$\\
\hline
$\hspace{1.5cm}T_{\nu13}/\rm GeV\hspace{1.5cm}$ & $\hspace{1.5cm}-500\hspace{1.5cm}$ &$\hspace{1.5cm}500\hspace{1.5cm}$\\
\hline
$\hspace{1.5cm}T_\nu/\rm GeV\hspace{1.5cm}$ & $\hspace{1.5cm}-1500\hspace{1.5cm}$ &$\hspace{1.5cm}1500\hspace{1.5cm}$\\
\hline
$\hspace{1.5cm}T_{e13}/\rm GeV\hspace{1.5cm}$ & $\hspace{1.5cm}-500\hspace{1.5cm}$ &$\hspace{1.5cm}500\hspace{1.5cm}$\\
\hline
$\hspace{1.5cm}T_e/\rm GeV\hspace{1.5cm}$ & $\hspace{1.5cm}-2500\hspace{1.5cm}$ &$\hspace{1.5cm}2500\hspace{1.5cm}$\\
\hline
$\hspace{1.5cm}M_{BB'}/\rm GeV\hspace{1.5cm}$ & $\hspace{1.5cm}100\hspace{1.5cm}$ &$\hspace{1.5cm}3000\hspace{1.5cm}$\\
\hline
$\hspace{1.5cm}M_{BL}/\rm GeV\hspace{1.5cm}$ & $\hspace{1.5cm}500\hspace{1.5cm}$ &$\hspace{1.5cm}5000\hspace{1.5cm}$\\
\hline
$\hspace{1.5cm}M^2_{\tilde{L}13}/\rm GeV^2\hspace{1.5cm}$ &$\hspace{1.5cm}0\hspace{1.5cm}$& $\hspace{1.5cm}10^5\hspace{1.5cm}$\\
\hline
$\hspace{1.5cm}M^2_{\tilde{L}}/\rm GeV^2\hspace{1.5cm}$ &$\hspace{1.5cm}1\times10^5\hspace{1.5cm}$& $\hspace{1.5cm}3\times10^6\hspace{1.5cm}$\\
\hline
$\hspace{1.5cm}M^2_{\tilde{E}13}/\rm GeV^2\hspace{1.5cm}$ &$\hspace{1.5cm}0\hspace{1.5cm}$& $\hspace{1.5cm}10^5\hspace{1.5cm}$\\
\hline
$\hspace{1.5cm}M^2_{\tilde{E}}/\rm GeV^2\hspace{1.5cm}$ &$\hspace{1.5cm}2\times10^{5}\hspace{1.5cm}$& $\hspace{1.5cm}3\times10^6\hspace{1.5cm}$\\
\hline
\end{tabular}
\end{table*}

Fig.\ref {tauepik24}(b) shows the change of BR($\tau \rightarrow e \pi^+K^-$) with the parameters $g_{YB}$ and $M_{\tilde{L}13}^2$. From the spatial distribution of points, when $g_{YB}$ takes a negative value, especially in the range of -0.4 to -0.1, and at the same time $M_{\tilde{L}13}^2$ is relatively large ($M_{\tilde{L}13}^2>6\times10^4~{\rm GeV^2}$), $\textcolor{red}{\bullet}$ are the most concentrated, which indicates that the corresponding branching ratio is highest in this region. Conversely, the lower right corner of the graph is mainly occupied by $\textcolor{blue}{\blacklozenge}$. In other words, the distribution of the low branching ratio is the most dense when $g_{YB}$ is close to the positive value (0 to 0.2) and $M_{\tilde{L}13}^2$ is smaller (less than $3\times10^4~{\rm GeV^2}$). $\textcolor{green}{\blacktriangle}$ are predominantly located in the middle of Fig.\ref {tauepik24}(b), roughly forming a transition zone from the lower left to the upper right. On the whole, the influence of $M_{\tilde{L}13}^2$ on the branching ratio is more direct, and the change of $g_{YB}$ also has an obvious secondary effect. With the increase of $M_{\tilde{L}13}^2$, the branching ratio gradually increases. Under the same $M_{\tilde{L}13}^2$, if $g_{YB}$ is smaller, $\textcolor{red}{\bullet}$ are also more likely to appear.

\begin{figure}[ht]
\setlength{\unitlength}{5mm}
\centering
\includegraphics[width=2.9in]{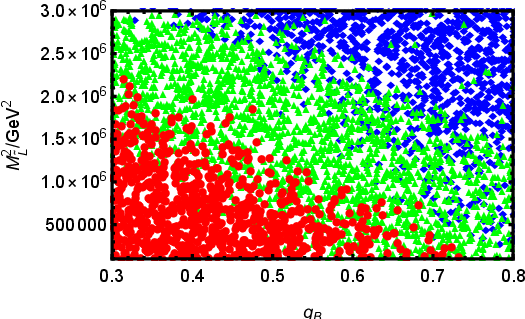}
\caption{Under the premises of current limits on LFV decays $\tau \rightarrow e \pi^+K^-$ and $\tau \rightarrow e \gamma$, reasonable parameter space is selected to scatter points. $\textcolor{blue}{\blacklozenge}~(0<\rm BR(\tau \rightarrow e \pi^+K^-)<4\times 10^{-11}),~\textcolor{green}{\blacktriangle}~(4\times 10^{-11}\leq \rm BR(\tau \rightarrow e \pi^+K^-)<8\times 10^{-11}),~ \textcolor{red}{\bullet}~(8\times 10^{-11}\leq \rm BR(\tau \rightarrow e \pi^+K^-)<3.7\times 10^{-8})$.}\label{tauepik3}
\end{figure}

Under the conditions $\lambda=0.4,~\tan\beta=25,~T_\nu=1~{\rm TeV},~T_e=1.5~{\rm TeV},~v_S=4~{\rm TeV},~M_{BB'}=0.1~{\rm TeV},~M_{BL}=1~{\rm TeV}$ and $M_{\tilde{L}13}^2=0.05~{\rm TeV^2}$, we plot $g_B$ varying with $M^2_{\tilde{L}}$ in Fig.\ref {tauepik3}. These parameter ranges are given in the Table \ref{c}. $\textcolor{red}{\bullet}$ are mainly concentrated in the lower left part of Fig.\ref {tauepik3}, when $g_B<0.5$ and $M^2_{\tilde{L}}<1.5\times10^6~{\rm GeV^2}$. The points with a high branching ratio are more densely distributed, indicating that the parameter combination corresponding to this area is more likely to enhance the decay rate of $\tau \rightarrow e \pi^+K^-$. $\textcolor{green}{\blacktriangle}$ are distributed in a strip in the middle of Fig.\ref {tauepik3}, with a wide horizontal extension, indicating that this is the intermediate region where the branching ratio transitions from high to low. $\textcolor{blue}{\blacklozenge}$ are distributed in large numbers in the upper right corner of Fig.\ref {tauepik3}. In the area of $g_B>0.6$ and $M^2_{\tilde{L}}>2\times10^6~{\rm GeV^2}$, $\textcolor{red}{\bullet}$ and $\textcolor{green}{\blacktriangle}$ are almost completely disappearing, and only $\textcolor{blue}{\blacklozenge}$ exist, indicating that this parameter interval has a strong inhibitory effect on LFV decay. The dividing line shows a clear diagonal distribution, and the three regions of red-green-blue basically transition from the bottom left to the top right. This shows that increasing either $g_B$ or $M^2_{\tilde{L}}$ alone can reduce BR($\tau \rightarrow e \pi^+K^-$), and if both are increased at the same time, the inhibitory effect is more obvious. This diagram clearly reveals the sensitivity of the $\tau \rightarrow e \pi^+K^-$ decay process to the two parameters $g_B$ and $M^2_{\tilde{L}}$, both of which play a key role in controlling the size of the branching ratio.

\begin{table*}
\caption{Scanning parameters for Fig.{\ref {tauepik3}}}\label{c}
\begin{tabular}{|c|c|c|}
\hline
Parameters&Min&Max\\
\hline
$\hspace{1.5cm}\lambda_2\hspace{1.5cm}$ &$\hspace{1.5cm}-0.3\hspace{1.5cm}$& $\hspace{1.5cm}-0.05\hspace{1.5cm}$\\
\hline
$\hspace{1.5cm}\kappa\hspace{1.5cm}$ &$\hspace{1.5cm}0.01\hspace{1.5cm}$& $\hspace{1.5cm}0.7\hspace{1.5cm}$\\
\hline
$\hspace{1.5cm}g_{YB}\hspace{1.5cm}$ &$\hspace{1.5cm}-0.4\hspace{1.5cm}$ &$\hspace{1.5cm}0.2\hspace{1.5cm}$\\
\hline
$g_B$ & $\hspace{1.5cm}0.3\hspace{1.5cm}$ &$\hspace{1.5cm}0.8\hspace{1.5cm}$\\
\hline
$\hspace{1.5cm}M^2_{\tilde{L}}/\rm GeV^2\hspace{1.5cm}$ &$\hspace{1.5cm}1\times10^5\hspace{1.5cm}$& $\hspace{1.5cm}3\times10^6\hspace{1.5cm}$\\
\hline
$\hspace{1.5cm}M^2_{\tilde{E}}/\rm GeV^2\hspace{1.5cm}$ &$\hspace{1.5cm}2\times10^{5}\hspace{1.5cm}$& $\hspace{1.5cm}3\times10^6\hspace{1.5cm}$\\
\hline
\end{tabular}
\end{table*}

\subsection{The process of $\tau \rightarrow e K^+K^-$}

\begin{figure}[ht]
\setlength{\unitlength}{5mm}
\centering
\includegraphics[width=2.9in]{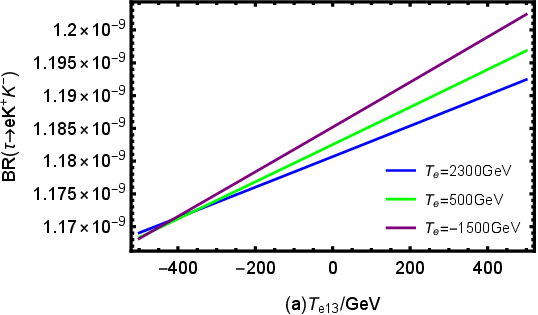}
\setlength{\unitlength}{5mm}
\centering
\includegraphics[width=2.9in]{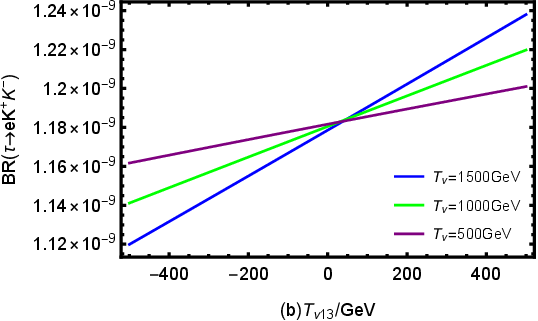}
\caption{The influence of various parameters on BR($\tau \rightarrow e K^+K^-$): In (a), $T_\nu=1~{\rm TeV}$. In (b), $T_e=1.5~{\rm TeV}$.}\label{tauekk}
\end{figure}

With the parameters $\kappa=0.1,~g_{YB}=0.1,~g_B=0.3,~\lambda=0.4,~\lambda_2=-0.25,~\tan\beta=25,~v_S=4~{\rm TeV},~M_{BB'}=0.1~{\rm TeV},~M_{BL}=1~{\rm TeV},~M_{\tilde{L}13}^2=0.05~{\rm TeV^2},~M_{\tilde{L}}^2=0.16~{\rm TeV^2},~M_{\tilde{E}}^2=1.7~{\rm TeV^2}$, we paint BR($\tau \rightarrow e K^+K^-$) schematic diagrams affected by different parameters in Fig.\ref{tauekk}.

The trend of BR($\tau \rightarrow e K^+K^-$) with $T_{e13}$ is investigated in Fig.\ref{tauekk}(a). Three lines are analyzed: a blue line represents $T_e=2300~{\rm GeV}$, a green line corresponds to $T_e=500~{\rm GeV}$, and a purple line indicates $T_e=-1500~{\rm GeV}$, demonstrating the sensitivity of the branching ratio to these parameters. All three curves show a monotonous upward trend. As $T_{e13}$ changes from negative to positive,the branching ratio increases. The purple line is always at the top, the green line is in the middle, and the blue line is at the bottom. That is to say, under the same $T_{e13}$ condition, the smaller the $T_e$, the larger the branching ratio. The spacing between the three line slightly expands as $T_{e13}$ increases. From a quantitative perspective, in the N-B-LSSM model, the mass square matrix of slepton is a $6\times6$ dimensional sturcture, which is jointly determined by the softbreaking mass term and the trilinear coupling term of the left-handed and right-handed three-generation lepton superparticles. The matrix can be split into $3\times3$ flavor sub-blocks, corresponding to the flavor-conserving and flavor-violating components respectively. The parameter $T_{e13}$ appears in the flavor-violating off-diagonal part of the matrix, while $T_e$ corresponds to the flavor-conserving diagonal element. The flavor mixing is introduced in the diagonalisation of slepton weak interaction eigenstates into the mass eigenstates, and the magnitude of the mixing angle is related to the ratio of off-diagonal element to a diagonal element $\frac{T_{e13}}{T_{e}}$. Therefore, the larger $T_{e13}$ is, the smaller $T_e$ is, and the BR($\tau \rightarrow e K^+K^-$) increases.

We study the effect of the parameter $T_{\nu13}$ on BR($\tau \rightarrow e K^+K^-$) using blue line ($T_\nu=1500~{\rm GeV}$), green line ($T_\nu=1000~{\rm GeV}$) and purple line ($T_\nu=500~{\rm GeV}$) in Fig.\ref{tauekk}(b). $T_{\nu13}$ represents the trilinear soft SUSY breaking term between the first and third generations in the CP-even and CP-odd sneutrino mass matrices, which is a lepton flavor-violating parameter. The branching ratio of the three curves increases with the increase of $T_{\nu13}$, and they intersect at $T_{\nu13}=0~{\rm GeV}$. As $T_{\nu13}$ moves away from the 0 point, the three curves gradually bifurcate. In the area of $T_{\nu13}<0~{\rm GeV}$, the purple line is higher than the green line, which is higher than the blue line. The smaller the $T_{\nu}$, the larger the branching ratio. In the region where $T_{\nu13}>0~{\rm GeV}$, the order is reversed. The blue line is the highest, followed by the green line, and the purple line is the lowest. The larger the $T_{\nu}$, the greater the branching ratio. This opposite sorting trend occurs because the positive or negative sign of $T_{\nu13}$ affects the sign of the interference term when CP-even and CP-odd sneutrinos mix, thereby influencing the specific form of flavor mixing and changing the dependence direction of the branching ratio on $T_{\nu}$.

It is worth noting that the amount of relative change in the vertical axis range of Fig.\ref{tauekk} is small. Therefore, it can be concluded that $T_e$,~$T_{e13}$,~$T_\nu$ and $T_{\nu13}$ do have an impact on BR($\tau \rightarrow e K^+K^-$), but it is relatively small. This small variation shows that they are not the main parameters that control the process.

\begin{figure}[ht]
\setlength{\unitlength}{5mm}
\centering
\includegraphics[width=2.7in]{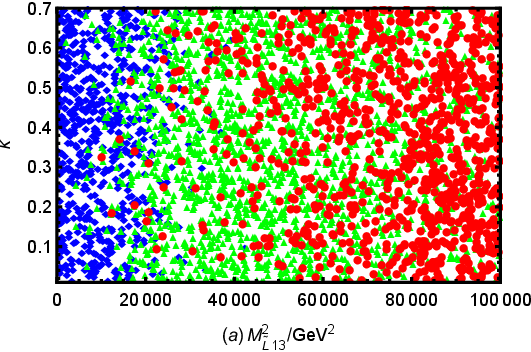}
\setlength{\unitlength}{5mm}
\centering
\includegraphics[width=2.9in]{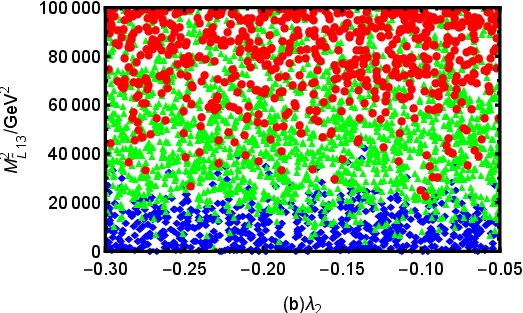}
\setlength{\unitlength}{5mm}
\centering
\includegraphics[width=2.9in]{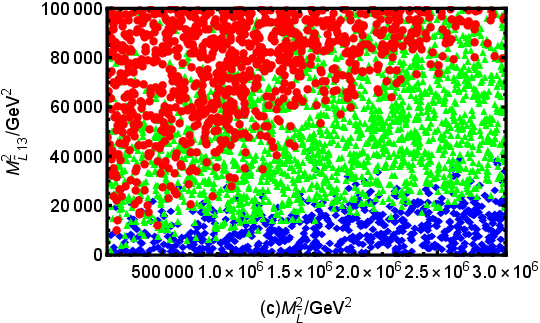}
\caption{Under the premises of current limits on LFV decays $\tau \rightarrow e K^+K^-$ and $\tau \rightarrow e \gamma$, reasonable parameter space is selected to scatter points. $\textcolor{blue}{\blacklozenge}~(0<\rm BR(\tau \rightarrow e K^+K^-)<3\times 10^{-11}),~\textcolor{green}{\blacktriangle}~(3\times 10^{-11}\leq \rm BR(\tau \rightarrow e K^+K^-)<5\times 10^{-10}),~ \textcolor{red}{\bullet}~(5\times 10^{-10}\leq \rm BR(\tau \rightarrow e K^+K^-)<3.4\times 10^{-8})$.}\label{tauekk1}
\end{figure}

In order to better study LFV and find a reasonable parameter space in the process of $\tau \rightarrow e K^+K^-$, we study the effects of parameters $\kappa$,~$\lambda_2$,~$M_{\tilde{L}13}^2$ and $M_{\tilde{L}}^2$, and draw the scatter diagrams of a certain parameter space in Fig.\ref{tauekk1}. We scatter points according to the parameters given in Table \ref{bd} to obtain Fig.\ref{tauekk1}(a)(b)(c).

In the study of the LFV process, the B-LSSM model has been widely investigated as an extension of the MSSM, and the additional $U(1)_{B-L}$ symmetry introduced has brought rich particle spectra and physical properties. The N-B-LSSM model used in this paper is structurally similar to the B-LSSM, but adds the new coupling terms and introduces  several new parameters. Besides the parameter $\lambda$ discussed in the previous subsection on $\tau \rightarrow e \pi^+\pi^-$, these parameters also include $\kappa$ and $\lambda_2$, which do not exist in the traditional B-LSSM framework. Therefore, it is of great significance to study the numerical impact of these new parameters on the LFV process.

Fig.\ref{tauekk1}(a) and Fig.\ref{tauekk1}(b) show the distribution of BR$(\tau \rightarrow e K^+K^-)$ on the parameter plane ($M_{\tilde{L}13}^2$,~$\kappa$) and ($\lambda_2$,~$M_{\tilde{L}13}^2$) respectively. Judging from the overall distribution of the two diagrams, within the range of the selected parameters, the branching ratio increases with the increase of $M_{\tilde{L}13}^2$. Whether in different processes or different diagrams, the increase of $M_{\tilde{L}13}^2$ always leads to the rise of the branching ratio, demonstrating its crucial role as a source of flavor mixing. This unified trend further supports its dominant position in the LFV processes. Both $\kappa$ (varying within the range of 0 to 0.7) and $\lambda_2$ (scanning between -0.3 and -0.05) have an even distribution of data points and do not show a clear trend. This suggests that BR$(\tau \rightarrow e K^+K^-)$ exhibits a moderate dependence on these two parameters, with the observed effects remaining relatively mild. $\kappa$ is the parameter in the term $\frac{1}{3}\kappa\hat{S}\hat{S}\hat{S}$ of the superpotential. $\kappa$ has relation with the Higgs tree level potential and Higgs mass matrix through the mixing with Higgs singlet $\hat{S}$. $\lambda_2$ emerges in the term $\lambda_2\hat{S}\hat{\chi}_1\hat{\chi}_2$ of the superpotential. Because $\hat{\chi}_1$ and $\hat{\chi}_2$ are Higgs singlets, the term including $\lambda_2$ give contributions to the CP-even Higgs mass squared matrix. The two parameters belong to the extended structure of the Higgs sector and not directly appear in the vertices or intermediate state propagrators of $\tau \rightarrow e K^+K^-$ process. Consequently, it is difficult to significantly alter the amplitude of the process by changing these parameters, resulting in the branching ratio is insensitive to them.

Fig.\ref{tauekk1}(c) illustrates the variation of BR$(\tau \rightarrow e K^+K^-)$ in the two-dimensional parameter plane of $M_{\tilde{L}}^2$ and $M_{\tilde{L}13}^2$. $\textcolor{red}{\bullet}$ are mainly located in the upper left part of the image, specifically in the area where $M_{\tilde{L}}^2$ is smaller and $M_{\tilde{L}13}^2$ is larger. In this region, the branching ratio is more likely to exceed $10^{-10}$. $\textcolor{green}{\blacktriangle}$ are predominantly distributed in the middle zone, forming a transition band along the direction from bottom left to top right, showing an obvious oblique structure of the colour boundary. $\textcolor{blue}{\blacklozenge}$ are clustered in the lower right corner area, indicating that the branching ratio is relatively small when $M_{\tilde{L}}^2$ is larger and $M_{\tilde{L}13}^2$ is smaller. As can be seen from
Fig.\ref{tauekk1}(c), the branching ratio is significantly positively correlated with $M_{\tilde{L}13}^2$. $M_{\tilde{L}}^2$ has a secondary but still important effect on BR$(\tau \rightarrow e K^+K^-)$, inhibiting the branching ratio.

\begin{table*}
\caption{Scanning parameters for Fig.{\ref {tauekkx1}}}
\begin{tabular}{|c|c|c|}
\hline
Parameters&Min&Max\\
\hline
$\hspace{1.5cm}T_{e13}/\rm GeV\hspace{1.5cm}$ & $\hspace{1.5cm}-500\hspace{1.5cm}$ &$\hspace{1.5cm}500\hspace{1.5cm}$\\
\hline
$\hspace{1.5cm}T_{\nu13}/\rm GeV\hspace{1.5cm}$ & $\hspace{1.5cm}-500\hspace{1.5cm}$ &$\hspace{1.5cm}500\hspace{1.5cm}$\\
\hline
\end{tabular}
\end{table*}

\begin{figure}[ht]
\setlength{\unitlength}{5mm}
\centering
\includegraphics[width=2.9in]{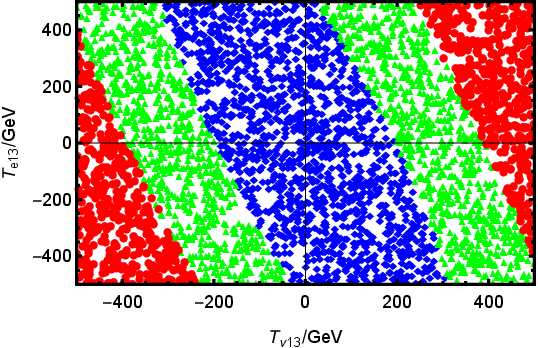}
\caption{Under the premises of current limits on LFV decays $\tau \rightarrow e K^+K^-$ and $\tau \rightarrow e \gamma$, reasonable parameter space is selected to scatter points. $\textcolor{blue}{\blacklozenge}~(0<\rm BR(\tau \rightarrow e K^+K^-)<5\times 10^{-14}),~\textcolor{green}{\blacktriangle}~(5\times 10^{-14}\leq \rm BR(\tau \rightarrow e K^+K^-)<2\times 10^{-13}),~ \textcolor{red}{\bullet}~(2\times 10^{-13}\leq \rm BR(\tau \rightarrow e K^+K^-)<3.4\times 10^{-8})$.}\label{tauekkx1}
\end{figure}

To further investigate the effects of flavor-violating parameters on the LFV decay $\tau \to e K^+K^-$, we perform a detailed analysis of the dependence of the branching ratio on the trilinear couplings $T_{e13}$ and $T_{\nu13}$. In Fig.\ref {tauekkx1}, the horizontal and vertical axes correspond to $T_{\nu13}$ and $T_{e13}$, respectively. The markers indicate the predicted branching ratio $\text{BR}(\tau \to e K^+K^-)$ under the current experimental bounds (including $\text{BR}(\tau \to e\gamma)$ and $\text{BR}(\tau \to e K^+K^-)$). For better visualization, the points are categorized into three regions: $\textcolor{blue}{\blacklozenge}~(0<\rm BR(\tau \rightarrow e K^+K^-)<5\times 10^{-14}),~\textcolor{green}{\blacktriangle}~(5\times 10^{-14}\leq \rm BR(\tau \rightarrow e K^+K^-)<2\times 10^{-13}),~ \textcolor{red}{\bullet}~(2\times 10^{-13}\leq \rm BR(\tau \rightarrow e K^+K^-)<3.4\times 10^{-8})$. The red region corresponds to relatively large branching ratios, though still below the experimental upper bound.

From this figure, several important features can be observed: In the region where $T_{\nu13} < 0$, for a fixed $T_{\nu13}$, increasing $T_{e13}$ significantly suppresses the branching ratio. Similarly, for a fixed $T_{e13}$, increasing $T_{\nu13}$ also leads to a decrease in $\text{BR}(\tau \rightarrow e K^+K^-)$. This implies that in this region, both parameters act to suppress the LFV signal. In the region where $T_{\nu13} > 0$, the behavior is opposite: increasing either $T_{e13}$ or $T_{\nu13}$ leads to a larger $\text{BR}(\tau \rightarrow e K^+K^-)$. This indicates a synergistic enhancement effect between the two parameters in this parameter space. The combination of these trends results in a clear diagonal pattern across the plot. Higher $\text{BR}(\tau \rightarrow e K^+K^-)$ values are mainly located toward the outer regions of the first and third quadrants, while lower $\text{BR}(\tau \rightarrow e K^+K^-)$ values concentrate near the middle region. This indicates that the influence of $T_{e13}$ and $T_{\nu13}$ not only depends on their absolute magnitudes but also significantly on their relative signs.

From a theoretical perspective, $T_{e13}$ arises from trilinear soft SUSY breaking terms in the charged slepton sector, while $T_{\nu13}$ affects the mass matrices of both CP-even and CP-odd sneutrinos. These parameters alter the flavor structure and mixing of sleptons and sneutrinos, thereby modulating the LFV transition amplitude. Depending on their values and signs, their contributions can interfere constructively or destructively, leading to the pattern observed in the figure.

Furthermore, although the dependence of $\text{BR}(\tau \to e K^+K^-)$ on $T_{e13}$ and $T_{\nu13}$ is clearly visible in the trend, the overall magnitude remains relatively small across most of the scanned parameter space. In particular, the majority of the parameter points yield $\text{BR}(\tau \rightarrow e K^+K^-)$ values in the range of $10^{-14}$ to $10^{-13}$, significantly below the experimental limit of $3.4 \times 10^{-8}$. Only a small subset of points (mainly near the edges of the first and third quadrants) reach higher values greater than $2 \times 10^{-13}$. This suggests that the influence of $T_{e13}$ and $T_{\nu13}$ on LFV is important, but reaching experimental sensitivity also needs contributions from other parameters.

\begin{table*}
\caption{Scanning parameters for Fig.{\ref {tauekkx2}}}
\begin{tabular}{|c|c|c|}
\hline
Parameters&Min&Max\\
\hline
$\hspace{1.5cm}M_{\tilde{L}13}/\rm GeV^2\hspace{1.5cm}$ & $\hspace{1.5cm}0\hspace{1.5cm}$ &$\hspace{1.5cm}10^4\hspace{1.5cm}$\\
\hline
$\hspace{1.5cm}T_{e13}/\rm GeV\hspace{1.5cm}$ & $\hspace{1.5cm}-500\hspace{1.5cm}$ &$\hspace{1.5cm}500\hspace{1.5cm}$\\
\hline
\end{tabular}
\end{table*}

\begin{figure}[ht]
\setlength{\unitlength}{5mm}
\centering
\includegraphics[width=2.9in]{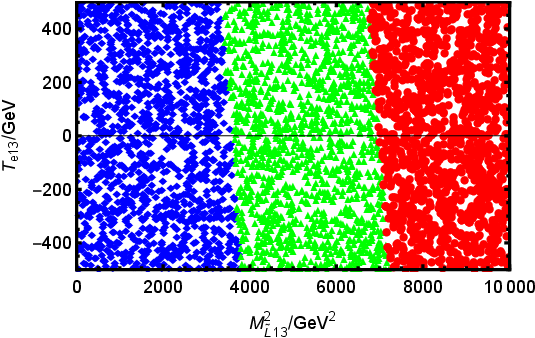}
\caption{Under the premises of current limits on LFV decays $\tau \rightarrow e K^+K^-$ and $\tau \rightarrow e \gamma$, reasonable parameter space is selected to scatter points. $\textcolor{blue}{\blacklozenge}~(0<\rm BR(\tau \rightarrow e K^+K^-)<6\times 10^{-12}),~\textcolor{green}{\blacktriangle}~(6\times 10^{-12}\leq \rm BR(\tau \rightarrow e K^+K^-)<2.3\times 10^{-11}),~ \textcolor{red}{\bullet}~(2.3\times 10^{-11}\leq \rm BR(\tau \rightarrow e K^+K^-)<3.4\times 10^{-8})$.}\label{tauekkx2}
\end{figure}

\begin{figure}[ht]
\setlength{\unitlength}{5mm}
\centering
\includegraphics[width=2.9in]{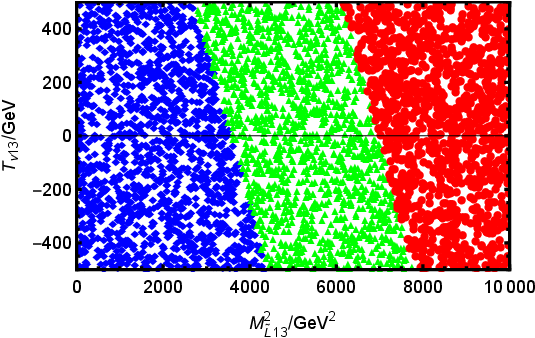}
\caption{Under the premises of current limits on LFV decays $\tau \rightarrow e K^+K^-$ and $\tau \rightarrow e \gamma$, reasonable parameter space is selected to scatter points. The BR($\tau \rightarrow e K^+K^-$) ranges marked by $\textcolor{blue}{\blacklozenge},~\textcolor{green}{\blacktriangle},~ \textcolor{red}{\bullet}$ are the same as in Fig.\ref {tauekkx2}.}\label{tauekkx3v}
\end{figure}

We include the contribution of $M^2_{\tilde{L}13}$ in Figs.\ref {tauekkx2},\ref{tauekkx3v} and perform a combined analysis with $T_{e13}$ and $T_{\nu13}$. As shown in the figures, $M^2_{\tilde{L}13}$ exhibits the most direct and significant impact on the branching ratio. As this parameter increases from 0 GeV$^2$ to 10000 GeV$^2$, the markers of the scattered points transition from $\textcolor{blue}{\blacklozenge}~(0<\rm BR(\tau \rightarrow e K^+K^-)<6\times 10^{-12})$ to $\textcolor{green}{\blacktriangle}~(6\times 10^{-12}\leq \rm BR(\tau \rightarrow e K^+K^-)<2.3\times 10^{-11})$, and eventually to $\textcolor{red}{\bullet} (2.3\times 10^{-11}\leq \rm BR(\tau \rightarrow e K^+K^-)<3.4\times 10^{-8})$. It indicates that $\text{BR}(\tau \to e K^+K^-)$ increases rapidly from the order of $10^{-12}$ to $10^{-11}$, approaching the current experimental limit. This behavior reflects the strong enhancement effect of $M^2_{\tilde{L}13}$ on LFV processes. In contrast, the effects of $T_{\nu13}$ and $T_{e13}$ are relatively mild. Within the same range of $M^2_{\tilde{L}13}$, variations in either $T_{\nu13}$ or $T_{e13}$ can lead to some changes in $\text{BR}(\tau \to e K^+K^-)$, but the amplitude and trend are significantly less pronounced than those induced by $M^2_{\tilde{L}13}$. In particular, the influence of $T_{\nu13}$ shows a more obvious directional tendency with a clearly sloped boundary, while the effect of $T_{e13}$ appears more gentle.

\begin{table*}
\caption{Scanning parameters for Fig.{\ref {tauekkx3v}}}
\begin{tabular}{|c|c|c|}
\hline
Parameters&Min&Max\\
\hline
$\hspace{1.5cm}M_{\tilde{L}13}/\rm GeV^2\hspace{1.5cm}$ & $\hspace{1.5cm}0\hspace{1.5cm}$ &$\hspace{1.5cm}10^4\hspace{1.5cm}$\\
\hline
$\hspace{1.5cm}T_{\nu13}/\rm GeV\hspace{1.5cm}$ & $\hspace{1.5cm}-500\hspace{1.5cm}$ &$\hspace{1.5cm}500\hspace{1.5cm}$\\
\hline
\end{tabular}
\end{table*}

This difference arises because $M^2_{\tilde{L}13}$ is the off-diagonal element in the left-handed slepton mass matrix, serving as a primary source of LFV. Since this parameter only appears in the mass term, its contribution to the branching ratio is monotonic and stable. Meanwhile, $T_{e13}$ and $T_{\nu13}$ act by introducing trilinear soft SUSY couplings between the third and first generations and modifying the off-diagonal structures of the slepton and sneutrino mass matrices, thereby affecting particle mixings and the LFV amplitudes. Their contributions are more indirect and nonlinear, leading to complex interference effects, yet generally limited in magnitude. These trends have already been analyzed in detail in Fig.\ref {tauekkx1}.

\section{discussion and conclusion}

In this paper, we study the LFV processes $\tau\rightarrow e M^+ M^-$ ($\tau \rightarrow e \pi^+\pi^-$, $\tau \rightarrow e \pi^+K^-$, $\tau \rightarrow e K^+K^-$) in the extended SUSY model N-B-LSSM. This model introduces right-handed neutrinos and three Higgs superfields $\hat{\chi}_1,~\hat{\chi}_2,~\hat{S}$ with the local gauge group $SU(3)_C\otimes SU(2)_L \otimes U(1)_Y\otimes U(1)_{B-L}$, so that the rotation matrices and interaction vertices in the N-B-LSSM are richer than those of MSSM. We construct the amplitude expressions of the corresponding processes based on the relevant Feynman diagrams.

Taking into account the upper limits on the branching ratio of $\tau\rightarrow e\gamma$, many diagrams of the numerical results are obtained after scanning large parameter spaces. The analysis of these numerical results indicates that $g_B$,~$g_{YB}$,~$\tan\beta$,~$M_{\tilde{L}}^2$,~$M_{\tilde{E}13}^2$ and $M_{\tilde{L}13}^2$ are sensitive parameters that have a significant impact on the branching ratios, $\lambda$,~$\lambda_2$,~$\kappa$,~$v_S$,~$T_\nu$,~$T_{\nu13}$,~$T_e$,~ $T_{e13}$,~$M_{BB'}$,~$M_{BL}$ and $M_{\tilde{E}}^2$ also affect the numerical results but not very large. In general, the non-diagonal elements corresponding to the initial and final leptons are the main sensitive parameters and LFV sources. We find that the order of magnitude of the $\tau\rightarrow e M^+ M^-$ branching ratios can reach approximately $10^{-11}-10^{-9}$. Most parameters can break the upper limit of the experiment and provide new ideas for finding NP.

\begin{acknowledgments}
This work is supported by National Natural Science Foundation of China (NNSFC)
(No.12075074), Natural Science Foundation of Hebei Province
(A2023201040, A2022201022, A2022201017, A2023201041), Natural Science Foundation of Hebei Education Department (QN2022173), Post-graduate's Innovation Fund Project of Hebei University (HBU2024SS042), the Project of the China Scholarship Council (CSC) No. 202408130113. X. Dong acknowledges support from Funda\c{c}\~{a}o para a Ci\^{e}ncia e a Tecnologia (FCT, Portugal) through the projects CFTP FCT Unit UIDB/00777/2020 and UIDP/00777/2020.
\end{acknowledgments}

\appendix
\section{The required Wilson coefficients}\label{A1}
In this section, we give out the concrete forms of the corresponding required Wilson coefficients as:
\begin{eqnarray}
&&C^{eq}_{VLR}=\sum_{F_1,F_2=\chi^\pm,\chi^\pm} \sum_{S_1,S_2=\tilde{\nu},\tilde{u}}-\frac{1}{2}I_9(x_{F_1},x_{F_2},x_{S_1},x_{S_2})H_L^{S_1\tau\bar{F_1}}H_R^{S_1 F_2\bar{e}}H_L^{S_2 F_1 \bar{d}}H_R^{S_2 d\bar{F_2}}\nonumber\\&&~~~~~~~~
-\sum_{F_1,F_2=\chi^0,\chi^0} \sum_{S_1,S_2=\tilde{e},\tilde{q}}\frac{1}{2}\Big[I_9(x_{F_1},x_{F_2},x_{S_1},x_{S_2})H_L^{S^*_1\tau\bar{F_1}}H_R^{S_1 F_2\bar{e}}H_L^{S_2 F_1 \bar{q}}H_R^{S^*_2 q\bar{F_2}}\nonumber\\&&~~~~~~~~-\frac{1}{2}I_8(x_{F_1},x_{F_2},x_{S_1},x_{S_2})H_L^{S^*_1\tau\bar{F_1}}H_R^{S_1 F_2\bar{e}}H_L^{S_2 F_2 \bar{q}}H_R^{S^*_2 q \bar{F_1 }}
\Big]\nonumber\\&&~~~~~~~~
+\sum_{F=\chi^0,\chi^{\pm}} \sum_{S=\tilde{e},\tilde{\nu}}\frac{1}{m_Z^{2}}\Big[\frac{1}{2}I_5(x_F,x_{S_1},x_{S_2})[H_R^{S_2 F \bar{e}}H_L^{S^* \tau \bar{F}}H^{Z S_1 S^*_1}H_R^{\bar{q} Z q}\nonumber\\&&~~~~~~~~
- H_R^{S F_2 \bar{e}}H_R^{Z F_1 \bar{F_2}}H_L^{S^* \tau \bar{F_1}}H_R^{\bar{q} Z q}]
+m_{F_1}m_{F_2}I_6(x_S,x_{F_2},x_{F_1})H_R^{S F_2 \bar{e}}H_L^{Z F_1 \bar{F_2}}H_L^{S^* \tau \bar{F_1}}H_R^{\bar{q} Z q}\Big]\nonumber\\&&~~~~~~~~
+\frac{1}{(m^2_e-m^2_\tau)}\Big[\frac{1}{2}I_1(x_F,x_S)(m^2_e-m^2_\tau ) [\frac{Q_q e^2}{k^2}H_R^{S^* F \bar{e}} H_L^{S \tau \bar{F}}\nonumber\\&&~~~~~~~~-\frac{1}{(k^2-m^2_Z)}H_R^{S^* F \bar{e}} H_L^{S \tau \bar{F}}H_L^{Z \tau \bar{e}}H_R^{\bar{q} Z q}]
+(m^2_e-m^2_\tau)[I_3(x_F,x_S)-I_4(x_F,x_S)]\nonumber\\&&~~~~~~~~\times[\frac{Q_q e^2}{k^2}m_F(m_e  H_L^{S^* F \bar{e}} H_L^{S \tau \bar{F}}+m_\tau H_R^{S^* F \bar{e}} H_R^{S \tau \bar{F}})-\frac{1}{(k^2-m^2_Z)}\nonumber\\&&~~~~~~~~\times m_F(m_e  H_L^{S^* F \bar{e}} H_L^{S \tau \bar{F}}H_L^{Z \tau \bar{e}}H_R^{\bar{q} Z q}+m_\tau H_R^{S^* F \bar{e}} H_R^{S \tau \bar{F}}H_L^{Z \tau \bar{e}}H_R^{\bar{q} Z q})]\Big]\nonumber\\&&~~~~~~~~
+\sum_{F=\chi^0} \sum_{S=\tilde{e}}\frac{-Q_q e^2}{k^2}\Big[\frac{1}{2}I_1(x_F,x_S) H_R^{S F \bar{e}} H_L^
{S^* \tau \bar{F}} +[I_2(x_F,x_S)-I_4(x_F,x_S)]\nonumber\\&&~~~~~~~~
\times[(m^2_\tau+m^2_e)H_R^{S F \bar{e}} H_L^{S^* \tau \bar{F}} +m_e m_\tau H_L^{S F \bar{e}} H_R^{S^* \tau \bar{F}} ]\nonumber\\&&~~~~~~~~+[I_2(x_F,x_S)-I_3(x_F,x_S)]
[m_F (m_e H_L^{S F \bar{e}} H_L^{S^* \tau \bar{F}} +m_\tau H_R^{S F \bar{e}} H_R^{S^* \tau \bar{F}} )]\Big]\nonumber\\&&~~~~~~~~
+\sum_{F=\chi^{\pm}} \sum_{S=\tilde{\nu}}\frac{Q_q e^2}{k^2}\Big[[\frac{1}{2}I_1(x_F,x_S)-{m^2_F} I_3(x_F,x_S)] H_R^{S F \bar{e}} H_L^{S^* \tau \bar{F}}\nonumber\\&&~~~~~~~~+[2I_4(x_F,x_S)-I_2(x_F,x_S)-I_3(x_F,x_S)]
[(m^2_\tau+m^2_e)H_R^{S F \bar{e}} H_L^{S^* \tau \bar{F}} \nonumber\\&&~~~~~~~~+m_e m_\tau H_L^{S F \bar{e}}  H_R^{S^* \tau \bar{F}} ]+[I_4(x_F,x_S)-I_3(x_F,x_S)]
[m_F (m_e H_L^{S F \bar{e}}  H_L^{S^* \tau \bar{F}} \nonumber\\&&~~~~~~~~+ m_\tau H_R^{S F \bar{e}}  H_R^{S^* \tau \bar{F}} )]\Big].
\end{eqnarray}

\begin{eqnarray}
&&C^{eq}_{SRR}=\sum_{F_1,F_2=\chi^\pm,\chi^\pm} \sum_{S_1,S_2=\tilde{\nu},\tilde{u}}\frac{1}{8}I_9(x_{F_1},x_{F_2},x_{S_1},x_{S_2})\Big[H_R^{S_1\tau\bar{F_1}}H_L^{S_1 F_2\bar{e}}H_R^{S_2 F_1 \bar{d}}H_L^{S_2 d\bar{F_2}}\nonumber\\&&~~~~~~~+H_L^{S_1\tau\bar{F_1}}H_R^{S_1 F_2\bar{e}}H_L^{S_2 F_1 \bar{d}}H_R^{S_2 d\bar{F_2}}+m_{F_1} m_{F_2}(H_L^{S_1\tau\bar{F_1}}H_L^{S_1 F_2\bar{e}}H_L^{S_2 F_1 \bar{d}}H_L^{S_2 d\bar{F_2}}\nonumber\\&&~~~~~~~-3H_R^{S_1\tau\bar{F_1}}H_R^{S_1 F_2\bar{e}}H_R^{S_2 F_1 \bar{d}}H_R^{S_2 d\bar{F_2}})\Big]\nonumber\\&&~~~~~~~
+\sum_{F_1,F_2=\chi^0,\chi^0} \sum_{S_1,S_2=\tilde{e},\tilde{q}}\frac{1}{8}I_9(x_{F_1},x_{F_2},x_{S_1},x_{S_2})\Big[H_R^{S^*_1\tau\bar{F_1}}H_L^{S_1 F_2\bar{e}}H_R^{S_2 F_1 \bar{q}}H_L^{S^*_2 q\bar{F_2}}\nonumber\\&&~~~~~~~+H_L^{S^*_1\tau\bar{F_1}}H_R^{S_1 F_2\bar{e}}H_L^{S_2 F_1 \bar{q}}H_R^{S^*_2 q\bar{F_2}}-H_L^{S^*_1\tau\bar{F_1}}H_L^{S_1 F_2\bar{e}}H_L^{S_2 F_2 \bar{q}}H_L^{S^*_2 q \bar{F_1 }}\nonumber\\&&~~~~~~~-H_R^{S^*_1\tau\bar{F_1}}H_R^{S_1 F_2\bar{e}}H_R^{S_2 F_2 \bar{q}}H_R^{S^*_2 q \bar{F_1 }}-H_R^{S^*_1\tau\bar{F_1}}H_L^{S_1 F_2\bar{e}}H_L^{S_2 F_2 \bar{q}}H_R^{S^*_2 q \bar{F_1 }}\nonumber\\&&~~~~~~~-H_L^{S^*_1\tau\bar{F_1}}H_R^{S_1 F_2\bar{e}}H_R^{S_2 F_2 \bar{q}}H_L^{S^*_2 q \bar{F_1 }}+m_{F_1} m_{F_2}(H_L^{S^*_1\tau\bar{F_1}}H_L^{S_1 F_2\bar{e}}H_L^{S_2 F_1 \bar{q}}H_L^{S^*_2 q\bar{F_2}}\nonumber\\&&~~~~~~~-3H_R^{S^*_1\tau\bar{F_1}}H_R^{S_1 F_2\bar{e}}H_R^{S_2 F_1 \bar{q}}H_R^{S^*_2 q\bar{F_2}}+4H_R^{S^*_1\tau\bar{F_1}}H_R^{S_1 F_2\bar{e}}H_R^{S_2 F_2 \bar{q}}H_R^{S^*_2 q \bar{F_1 }})\Big].
\end{eqnarray}

\begin{eqnarray}
&&C^{eq}_{SRL}=\sum_{F_1,F_2=\chi^{\pm},\chi^\pm} \sum_{S_1,S_2=\tilde{\nu},\tilde{u}}-\frac{1}{2}I_8(x_{F_1},x_{F_2},x_{S_1},x_{S_2})H_R^{S_1\tau\bar{F_1}}H_R^{S_1 F_2\bar{e}}H_L^{S_2 F_1 \bar{d}}H_L^{S_2 d\bar{F_2}}\nonumber\\&&~~~~~~~-\frac{1}{8}I_9(x_{F_1},x_{F_2},x_{S_1},x_{S_2})\Big[m_{F_1}m_{F_2}(H_L^{S_1\tau\bar{F_1}}H_L^{S_1 F_2\bar{e}}H_L^{S_2 F_1 \bar{d}}H_L^{S_2 d\bar{F_2}}\nonumber\\&&~~~~~~~+H_R^{S_1\tau\bar{F_1}}H_R^{S_1 F_2\bar{e}}H_R^{S_2 F_1 \bar{d}}H_R^{S_2 d\bar{F_2}})+H_R^{S_1\tau\bar{F_1}}H_L^{S_1 F_2\bar{e}}H_R^{S_2 F_1 \bar{d}}H_L^{S_2 d\bar{F_2}}\nonumber\\&&~~~~~~~+H_L^{S_1\tau\bar{F_1}}H_R^{S_1 F_2\bar{e}}H_L^{S_2 F_1 \bar{d}}H_R^{S_2 d\bar{F_2}}\Big]\nonumber\\&&~~~~~~~
-\sum_{F_1,F_2=\chi^0,\chi^0} \sum_{S_1,S_2=\tilde{e},\tilde{q}}\frac{1}{2}I_8(x_{F_1},x_{F_2},x_{S_1},x_{S_2})
\Big[H_R^{S^*_1\tau\bar{F_1}}H_R^{S_1 F_2\bar{e}}H_L^{S_2 F_2 \bar{q}}H_L^{S^*_2 q \bar{F_1 }}\nonumber\\&&~~~~~~~+H_R^{S^*_1\tau\bar{F_1}}H_R^{S_1 F_2\bar{e}}H_L^{S_2 F_1 \bar{q}}H_L^{S^*_2 q\bar{F_2}}\Big]+\frac{1}{8}I_9(x_{F_1},x_{F_2},x_{S_1},x_{S_2})\nonumber\\&&~~~~~~~\times\Big[H_L^{S^*_1\tau\bar{F_1}}H_L^{S_1 F_2\bar{e}}H_L^{S_2 F_2 \bar{q}}H_L^{S^*_2 q \bar{F_1 }}+H_R^{S^*_1\tau\bar{F_1}}H_R^{S_1 F_2\bar{e}}H_R^{S_2 F_2 \bar{q}}H_R^{S^*_2 q \bar{F_1 }}\nonumber\\&&~~~~~~~+H_R^{S^*_1\tau\bar{F_1}}H_L^{S_1 F_2\bar{e}}H_L^{S_2 F_2 \bar{q}}H_R^{S^*_2 q \bar{F_1 }}+H_L^{S^*_1\tau\bar{F_1}}H_R^{S_1 F_2\bar{e}}H_R^{S_2 F_2 \bar{q}}H_L^{S^*_2 q \bar{F_1 }}\nonumber\\&&~~~~~~~-[m_{F_1}m_{F_2}(H_L^{S^*_1\tau\bar{F_1}}H_L^{S_1 F_2\bar{e}}H_L^{S_2 F_1 \bar{q}}H_L^{S^*_2 q\bar{F_2}}+H_R^{S^*_1\tau\bar{F_1}}H_R^{S_1 F_2\bar{e}}H_R^{S_2 F_1 \bar{q}}H_R^{S^*_2 q\bar{F_2}})\nonumber\\&&~~~~~~~+H_R^{S^*_1\tau\bar{F_1}}H_L^{S_1 F_2\bar{e}}H_R^{S_2 F_1 \bar{q}}H_L^{S^*_2 q\bar{F_2}}+H_L^{S^*_1\tau\bar{F_1}}H_R^{S_1 F_2\bar{e}}H_L^{S_2 F_1 \bar{q}}H_R^{S^*_2 q\bar{F_2}}]\Big].
\end{eqnarray}

\begin{eqnarray}
&&C^{eq}_{TRR}=\sum_{F_1,F_2=\chi^{\pm},\chi^\pm} \sum_{S_1,S_2=\tilde{\nu},\tilde{u}}-\frac{m_{F_1}m_{F_2}}{8}I_9(x_{F_1},x_{F_2},x_{S_1},x_{S_2})H_R^{S_1\tau\bar{F_1}}H_R^{S_1 F_2\bar{e}}H_R^{S_2 F_1 \bar{d}}H_R^{S_2 d\bar{F_2}}\nonumber\\&&~~~~~~~
-\sum_{F_1,F_2=\chi^0,\chi^0} \sum_{S_1,S_2=\tilde{e},\tilde{q}}\frac{1}{8}I_9(x_{F_1},x_{F_2},x_{S_1},x_{S_2})
\Big[H_R^{S^*_1\tau\bar{F_1}}H_R^{S_1 F_2\bar{e}}H_R^{S_2 F_2 \bar{q}}H_R^{S^*_2 q \bar{F_1 }}\nonumber\\&&~~~~~~~
+m_{F_1}m_{F_2}H_R^{S^*_1\tau\bar{F_1}}H_R^{S_1 F_2\bar{e}}H_R^{S_2 F_1 \bar{q}}H_R^{S^*_2 q\bar{F_2}}\Big].
\end{eqnarray}

\end{document}